\pdfoutput=1
\documentclass[10pt, conference, compsocconf]{template/IEEEtran}

\usepackage[table]{xcolor}
\usepackage{color}

\usepackage{cite}
\ifCLASSINFOpdf
\else
\fi
\usepackage[cmex10]{amsmath}
\usepackage{algpseudocode}
\usepackage[]{algorithm2e}

\usepackage{array}
\usepackage{fixltx2e}
\usepackage{ctable} % for \specialrule command
\usepackage{multirow}
\usepackage{multicol}
\usepackage{eqparbox}
\usepackage[font=footnotesize,caption=false]{subfig}

\usepackage{float}
\usepackage{hyperref}
\usepackage{epsfig}
\usepackage{enumerate}

\newtheorem{definition}{Definition}

\renewcommand{\IEEEQED}{\IEEEQEDopen}
\newcommand\Tstrut{\rule{0pt}{2ex}}       % "top" strut
\newcommand\Bstrut{\rule[-0.6ex]{0pt}{0pt}} % "bottom" strut
\newcommand{\TBstrut}{\Tstrut\Bstrut} % top&bottom struts
\definecolor{LightCyan}{rgb}{0.88,1,1}
\definecolor{DarkCyan}{rgb}{0,0.3,0.9}
\definecolor{LightGray}{gray}{0.85}
\algrenewcommand\alglinenumber[1]{\tiny #1:}

\usepackage{amssymb}
\usepackage{mathtools}
\usepackage{verbatim}
\usepackage{microtype}
\usepackage{changepage}
\usepackage{booktabs,fixltx2e}
\usepackage[flushleft]{threeparttable}

\SetCommentSty{mycommfont}
\usepackage[]{color}

\newcommand{\taskSet}{\ensuremath{T}}
\newcommand{\activitySet}{\ensuremath{A}}
\newcommand{\messageSet}{\ensuremath{M}}
\newcommand{\resourceSet}{\ensuremath{U}}
\newcommand{\activity}{\ensuremath{a}}
\newcommand{\resource}{\ensuremath{u}}
\newcommand{\numResources}{\ensuremath{m}}

\newcommand{\nActivities}{\ensuremath{n}}
\newcommand{\nJobs}{\ensuremath{n_{jobs}}}
\newcommand{\adjMatrix}{\ensuremath{A_G}}
\newcommand{\precedenceGraphMatrix}{\ensuremath{g}}

\newcommand{\period}{\ensuremath{p}}
\newcommand{\periodSet}{\ensuremath{P}}
\newcommand{\executionTime}{\ensuremath{e}}
\newcommand{\executionTimeSet}{\ensuremath{E}}
\newcommand{\jitter}{\ensuremath{jit}}
\newcommand{\promotedJitter}{\ensuremath{jit^{inher}}}

\newcommand{\lowerBound}[1]{\ensuremath{\hat{#1}}}

\newcommand{\naturalNumberSet}{\ensuremath{\mathbb{N}}}
\newcommand{\integerNumberSet}{\ensuremath{\mathbb{Z}}}
\newcommand{\rationalNumberSet}{\ensuremath{\mathbb{R}}}

\newcommand{\mapping}{\ensuremath{map}}
\newcommand{\mapVariable}{\ensuremath{q}}
\newcommand{\loadBalance}{\ensuremath{b}}
\newcommand{\utilization}{\ensuremath{r}}

\newcommand{\hyperPeriod}{\ensuremath{H}}

\newcommand{\scheduleOrderStart}{\ensuremath{s}}
\newcommand{\scheduleOrder}{\ensuremath{x}}
\newcommand{\numActivityOccurrence}{\ensuremath{n}}
\newcommand{\criticalTimeBefore}{\ensuremath{t^b}}
\newcommand{\criticalTimeAfter}{\ensuremath{t^a}}
\newcommand{\lengthOfSchedInterval}{\ensuremath{I}}

\newcommand{\ScheduledSet}{\ensuremath{Sch}}
\newcommand{\RootProblems}{\ensuremath{R}}
\newcommand{\activityToUnschedule}{\ensuremath{a_u}}
\newcommand{\activityToSchedule}{\ensuremath{a_c}}
\newcommand{\ScheduledFromScratch}{\ensuremath{Scratch}}
\newcommand{\scheduleHeur}{\ensuremath{S}}
\newcommand{\queueToSchedule}{\ensuremath{Q}}

\renewcommand{\baselinestretch}{0.93}

\SetAlFnt{\small}

\IEEEoverridecommandlockouts
\IEEEpubid{\makebox[\columnwidth]{ 0018-9340~
\copyright~2017 IEEE \hfill} \hspace{\columnsep}\makebox[\columnwidth]{}} 

\begin{document}
%
% article title
%\title{Efficient Configuration of Time-Division Multiplexed Memory Controllers}
%\title{Efficient Configuration of TDM Scheduled Memory Controllers}

\title{Time-Triggered Co-Scheduling of Computation and Communication with Jitter Requirements} %\thanks{This article is published in IEEE Transactions on Computers journal and can be found at \url{http://ieeexplore.ieee.org/document/7967685/\#full-text-section}}}

\author{\IEEEauthorblockN{Anna Minaeva$^{1}$, Benny Akesson$^{2}$, Zden{\v e}k Hanz{\' a}lek$^{1}$, Dakshina Dasari$^{3}$}
\IEEEauthorblockA{$^{1}$Faculty of Electrical Engineering and Czech Institute of Informatics, Robotics and Cybernetics, Czech Technical University in Prague}
\IEEEauthorblockA{$^{2}$CISTER/INESC TEC and ISEP}
\IEEEauthorblockA{$^{3}$Corporate Research, Robert Bosch GmbH, Germany}
}

\maketitle

\begin{abstract}

The complexity of embedded application design is increasing with growing user demands. In particular, automotive embedded systems are highly complex in nature, and their functionality is realized by a set of periodic tasks. These tasks may have hard real-time requirements and communicate over an interconnect. The problem is to efficiently co-schedule task execution on cores and message transmission on the interconnect so that timing constraints are satisfied. Contemporary works typically deal with zero-jitter scheduling, which results in lower resource utilization, but has lower memory requirements. This article focuses on jitter-constrained scheduling that puts constraints on the tasks jitter, increasing schedulability over zero-jitter scheduling. 
%and memory requirements on the final system as well.

The contributions of this article are: 1)~Integer Linear Programming and Satisfiability Modulo Theory model exploiting problem-specific information to reduce the formulations complexity to schedule small applications. 2)~A heuristic approach, employing three levels of scheduling scaling to real-world use-cases with~10000 tasks and messages. 3)~An experimental evaluation of the proposed approaches on a case-study and on synthetic data sets showing the efficiency of both zero-jitter and jitter-constrained scheduling. 
It shows that up to 28\% higher resource utilization can be achieved by having up to 10 times longer computation time with relaxed jitter requirements.

\vspace{1em}
\noindent
\emph{Cite as}: Anna Minaeva, Benny Akesson, Zden{\v e}k Hanz{\' a}lek, Dakshina Dasari, Time-Triggered Co-Scheduling of Computation and Communication with Jitter Requirements, \emph{IEEE Transactions on Computers}, 0018-9340, 10.1109/TC.2017.2722443.

\vspace{1em}
\noindent
\emph{Source code}: 
\url{https://github.com/CTU-IIG/CC_Scheduling_WithJitter}

\end{abstract}
\vspace{3mm}



%
%  Use this command to print the description
%
%\printccsdesc

% We no longer use \terms command

%\keywords{ACM proceedings; \LaTeX; text tagging}

\footnotetext{\copyright 2017. This manuscript version is made available under the CC-BY-NC-ND 4.0 license \url{http://creativecommons.org/licenses/by-nc-nd/4.0/}.
\\
This article is published in IEEE Transactions on Computers.
}

%%%%%%%%%%%%%%%%%%%%%%%%%%%%%%%%%%%%%%%%%%%%%%%%%%%%%%%%%%%%
\section{Introduction}
\label{introduction}

The complexity of embedded application design is increasing as a multitude of functionalities is incorporated to address growing user demands. The problem of \emph{non-preemptive} \emph{co-scheduling} of these applications on multiple cores and their communication via an interconnect can be found in automotive~\cite{tuohy2015intra,Bosch2004},
%,steffen2008design} ,navet2005trends
 avionics~\cite{honeywellAerospace} and other industries. For instance, automotive embedded systems (e.g. contemporary advanced engine control modules) are highly complex in nature, and their functionality is realized by a set of tightly coupled periodic tasks with \emph{hard real-time requirements} that communicate with each other over an interconnect. 
These tasks may be activated at different rates and execute sensing, control and actuation functions. %An example of a task set in the automotive domain is a cruise control application, where the vehicle speed is read by a speed sensor (sensing), is processed (control), and the throttle position is adjusted (actuation) to maintain the required speed (in a closed loop feedback mode). 
 Additionally, these embedded applications are required to realize many end-to-end control functions within predefined time bounds, while also executing the constituent tasks in a specific order. To reduce the cost of the resulting system, it is necessary to allocate resources efficiently.

\begin{figure*}[ht]%
    \centering
    \subfloat[Problem description]{{\includegraphics[width=6cm]{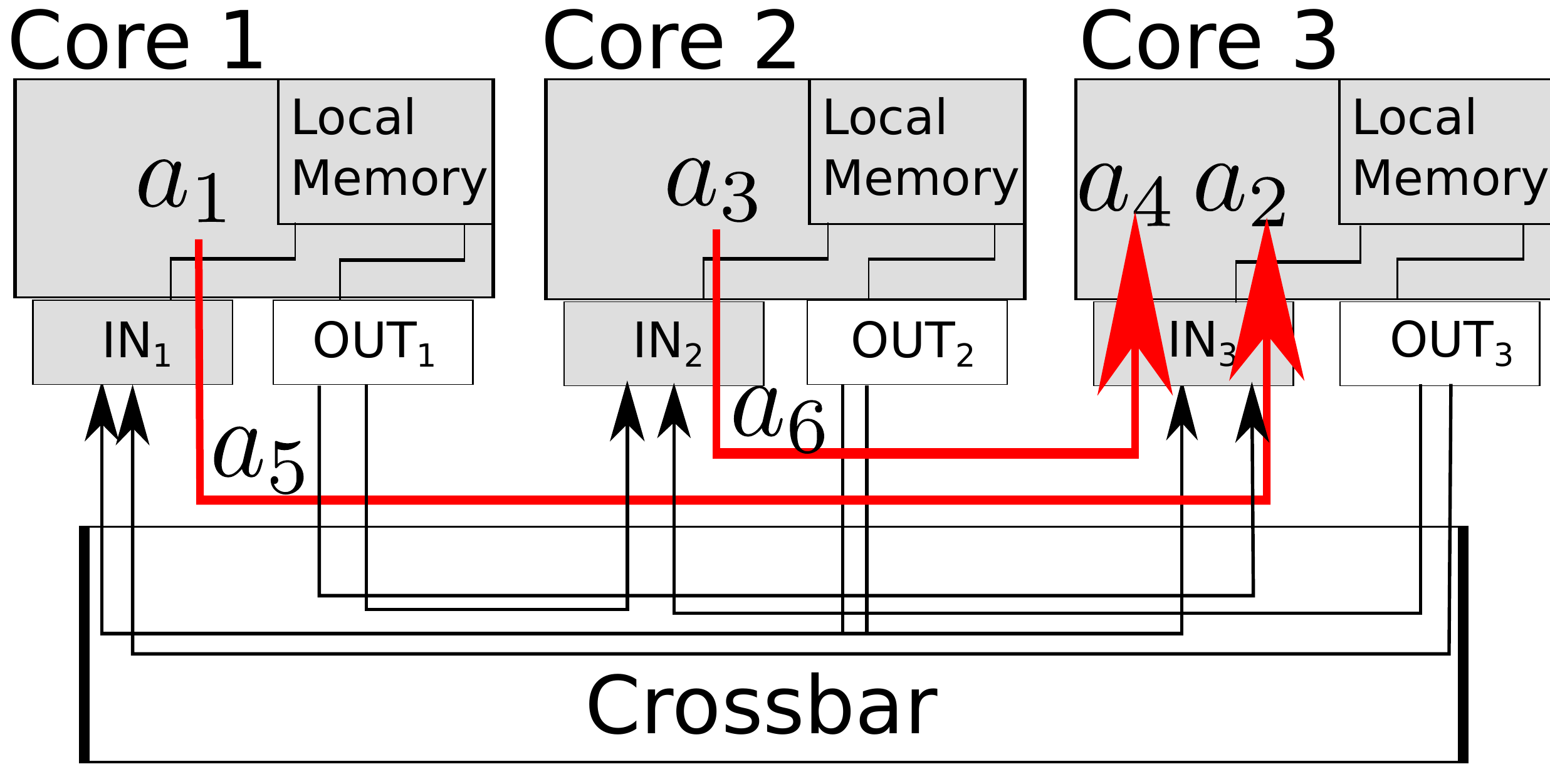} {\label{fig:problem}}}}%
    \qquad
    \subfloat[Schedule with jitter-constrained]{{\includegraphics[width=5cm]{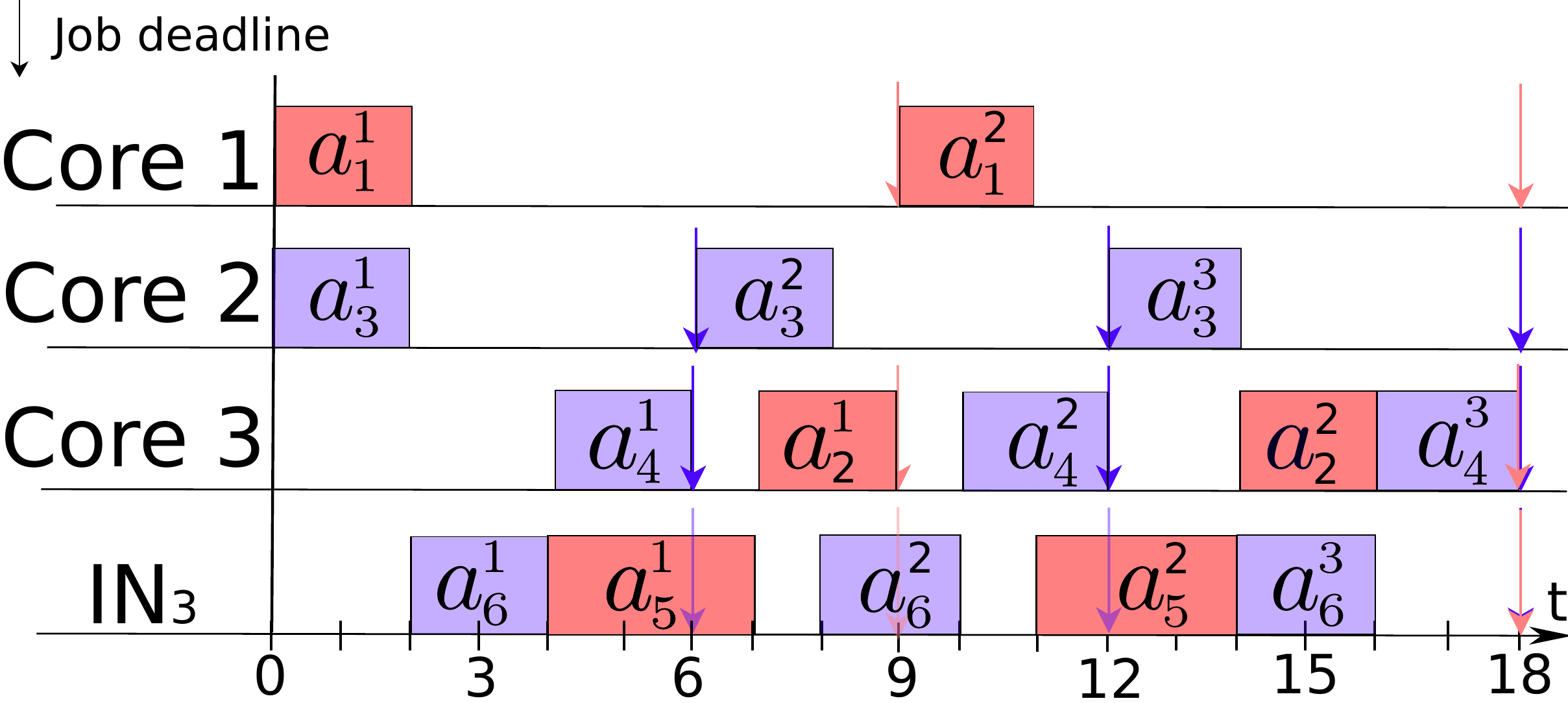}  {\label{fig:exampleNonZero}}}}%
    \qquad
    \subfloat[Schedule with zero-jitter]{{\includegraphics[width=5cm]{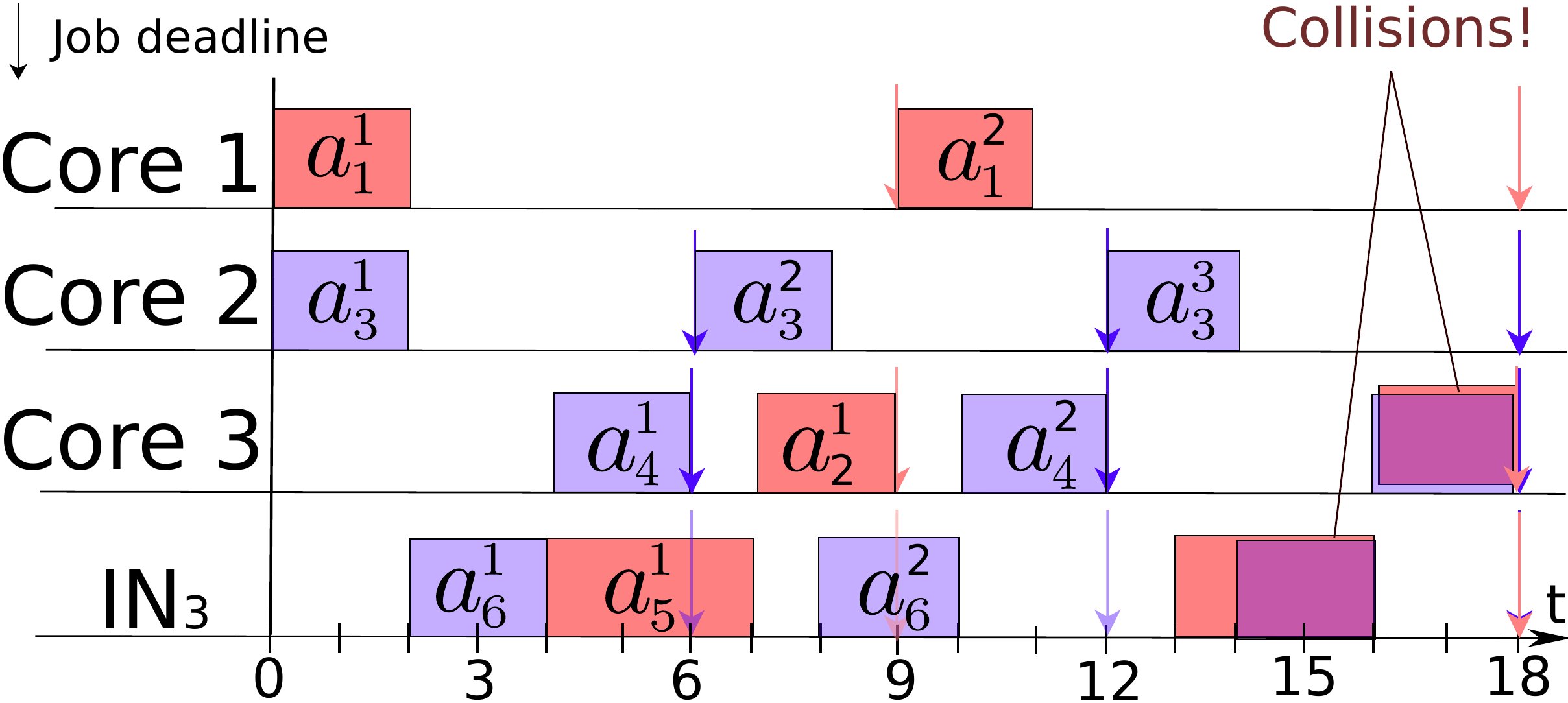}  {\label{fig:exampleZero}}}}%
    \caption{Multi-periodic scheduling problem description with examples of ZJ and JC solution, where  $\activity_5$ is a message between $\activity_1$ and $\activity_2$ and $\activity_6$ is a message between $\activity_3$ and $\activity_4$.}%
    \label{fig:example1}%
\end{figure*}

The considered problem is illustrated in Figure~\ref{fig:problem}, where tasks $\activity_1, \activity_2, \activity_3, \activity_4$ are mapped to Cores~1~to~3, where each core has its local memory and communicate via a crossbar switch. This architecture is inspired by Infineon AURIX TriCore~\cite{TriCore}. The crossbar switch is assumed to be a point-to-point connection that links an output port of each core with input ports of the remaining cores. Although there is \emph{no contention on output ports} since tasks on cores are statically scheduled, \emph{scheduling of the incoming messages on the input ports} must be done to prevent contention. Moreover, there are two chains of dependencies, indicated by thicker (red) arrows, i.e. $\activity_1 \rightarrow \activity_5 \rightarrow \activity_2$ and $\activity_3 \rightarrow \activity_6 \rightarrow \activity_4$. Note that although this example contains 6 resources to be scheduled, the only input port that must be scheduled in this case is the one of Core 3, since there are no incoming messages to other cores. 

The time-triggered approach, where the schedule is computed offline and repeated during execution is commonly used in scheduling safety-critical systems. However, contemporary time-triggered works mostly consider \emph{zero-jitter (ZJ) scheduling}~\cite{Steiner,craciunas2015combined} also called strictly periodic scheduling,
%,goswami2012time,Lukasie2012,zhang2014task},Biewer
 where the start time of an \emph{activity}, (i.e., task or message) is at a fixed offset (instant) in every period. If there are two consecutive periods in which the activity is scheduled at different times (relative to the period), we call it \emph{jitter-constrained (JC) scheduling}. On one hand, ZJ scheduling results in lower memory requirements,
% on each core of electronic control unit (ECU) 
since the schedule takes less space to store and typically needs less time to find an optimal solution. On the other hand, it puts too strict requirements on a schedule causing many problem instances to be infeasible, as we later show in Section~\ref{experiment}. This may lead to increasing requirements on the number of employed cores for the given application, and thus makes the system more expensive. Even though some applications or even systems are restricted to being ZJ, e.g. some systems in the avionics domain~\cite{al2012strictly}, many systems in the automotive domain allow JC scheduling~\cite{furst2009autosar, puffitsch2015off}. Therefore, this article explores the trade-off between JC and ZJ scheduling. Although not all activities have ZJ requirements, some of them are typically sensitive to the delay between consecutive occurrences, since it greatly influences the quality of control~\cite{di2000scheduling}. %,baruah1999scheduling,lin1996jitter
Assuming \emph{constrained jitter} instead of ZJ scheduling allows the resulting schedule to both satisfy strict jitter requirements of the jitter-critical activities and to have more freedom to schedule their non-jitter-critical counterpart. 

An example of the JC schedule for the problem in Figure~\ref{fig:problem} assumes that activities $\activity_1$, $\activity_2$ and $\activity_5$ have a required period of 9 time units, while $\activity_3$, $\activity_4$, and $\activity_6$ must be scheduled with a period of 6. The resulting JC schedule is shown in Figure~\ref{fig:exampleNonZero} with a hyper-period (length) of 18 time units, which is the least common multiple of both periods. Hence, activities $\activity_1$, $\activity_2$ and $\activity_5$ are scheduled 2 times and activities $\activity_3$, $\activity_4$, and $\activity_6$ are scheduled 3 times during one hyper-period, defining its number of \emph{jobs}, i.e. activity occurrences. Note that activities $\activity_2$ and $\activity_5$ are not scheduled with zero-jitter since $\activity_2$ in the first period is scheduled at time 7, while in the second period at time 5~(+9). Similarly, $\activity_5$ is scheduled at different times in the first and second periods (4 and 2(+9), respectively). In contrast, Figure~\ref{fig:exampleZero} illustrates that using ZJ scheduling results in collisions between $\activity_2$ and $\activity_4$ on Core 3, and between $\activity_5$ and $\activity_6$ in the crossbar switch. Moreover, an exact approach (see SMT formulation in Section~\ref{ILP}) can prove that the instance is infeasible with ZJ scheduling. Thus, if an application can tolerate some jitter in the execution of activities $\activity_2$ and $\activity_5$ without unacceptable quality degradation of control, then the system resources could be utilized more efficiently, as shown in Figure~\ref{fig:exampleNonZero}.

The three main contributions of this article are: 1)~Two models, one Integer Linear Programming (ILP) formulation and one Satisfiability Modulo Theory (SMT) model with problem-specific improvements to reduce the complexity and the computation time of the formulations. The two models are proposed due to significantly different computation times on problem instances of low and high complexity, respectively. The formulations optimally solve the problem for smaller applications with up to 50 activities in reasonable time. 2)~A heuristic approach, called 3-LS, employing a three-step approach that scales to real-world use-cases with more than 10000 activities. 3)~An experimental evaluation of the proposed solution for different jitter requirements on a synthetic data sets that quantifies the computation time and resource utilization trade-off and shows that relaxing jitter constraints allows to achieve on average up to 28\% higher resource utilization for the price of up to 10 times longer computation time. Moreover, the 3-LS heuristic is demonstrated on a case study of an engine management system, which it successfully solves in 43 minutes.
%It includes justification of reasonable memory requirements on each resource based on the statistics of real-world automotive benchmarks~\cite{WATERSAutomotive2015}. 
%Moreover, the use-case, inspired by real-world automotive benchmarks is used to validate the proposed approaches.

The rest of this article is organized as follows: the related work is discussed in Section~\ref{related_work}. Section~\ref{description} proceeds by presenting the system model and the problem formulation. The description of the ILP and SMT formulations and their computation time improvements follow in Section~\ref{ILP}. Section~\ref{heuristic} introduces the proposed heuristic approach for scheduling periodic activities, and Section~\ref{experiment} proceeds by presenting the experimental evaluation before concluding the article in Section~\ref{conclusions}.

%%%%%%%%%%%%%%%%%%%%%%%%%%%%%%%%%%%%%%%%%%%%%%%%%%%%%%%%%%%%%
\section{Related Work}
\label{related_work}
There are two approaches to solve the periodic scheduling problem with hard real-time requirements: 1)~\textit{Event-Triggered (ET) Scheduling}~\cite{Davis:2011:SHR:1978802.1978814}, where scheduling is performed during run-time of a system, triggered by events, and 2) The~\textit{Time-Triggered (TT) Scheduling} that builds schedules offline that are provably correct by construction. The TT scheduling is commonly adopted in safety-critical systems, due to the highly predictable behavior of the scheduled activities, simplifying design and verification~\cite{Kopetz2003}.

Even though this article targets the TT approach, the survey of related work would not be complete without mentioning articles that consider the ET paradigm. A broad survey of works related to periodic (hard real-time) scheduling is provided by Davis and Burns in~\cite{Davis:2011:SHR:1978802.1978814}. Next, Baruah et al. ~\cite{Baruah1996} introduce the notion of Pfair schedules, which relates to the concept of ZJ scheduling while scheduling preemptively, i.e. where execution of an activity can be preempted by another activity. Similarly to the ZJ approach that requires the time intervals equal to execution times of activities to be scheduled equidistantly in consecutive periods as a whole, Pfair requires equidistant allocation, while scheduling by intervals of one time instant. On the non-preemptive scheduling front, Jeffay at al. \cite{Jeffay} propose an approach to schedule periodic activities on a single resource with precedence constraints. The problem of co-scheduling tasks and messages in an event-triggered manner is considered in~\cite{tindell1994holistic, jayachandran2009end,kang2000parametric}. However, these works do not consider jitter constraints, as done in this article.

%In this paper, we assume that each activity is required to be scheduled exactly once in each time period, thus having absolute release times and deadlines on jobs of each activity in each time period. In contrast, \cite{Jeffay} works with jobs having relative release times and deadlines, where each job has constraints on a maximum delay, related to the previous job of the activity. 

% Moreover, Tindell and Clark in~\cite{tindell1994holistic} propose a holistic schedulability analysis that deals with scheduling both tasks and messages. Besides, Jayachandran and Abdelzaher in~\cite{jayachandran2009end} propose a way how to reduce multiprocessor schedulability analysis to a uniprocessor one. 

The TT approach attracted the attention of many researchers over the past twenty years for solving the problem of periodic scheduling. The pinwheel scheduling problem~\cite{pinwheel} can be viewed as a relaxation of the jitter-bounded scheduling concept, where each activity is required to be scheduled at least once during each predefined number of consecutive time units. 
%Similarly to~\cite{Jeffay}, the deadlines and release times of jobs, i.e. activity occurrences in each period are not absolute in the pinwheel problem, while 
If minimizing number of jobs, the solution of the pinwheel problem approaches the ZJ scheduling solution, since it tends to have an equidistant schedule for each activity. Moreover, the Periodic Maintenance Scheduling Problem~\cite{Bar-Noy2002} is identical to ZJ scheduling, as it requires jobs to be executed exactly a predefined number of time units apart.

Considering works that formulate the problem similarly, some authors deal with scheduling only one core~\cite{giannopoulou2015mixed, monot2012multisource}, while others focus only on interconnects~\cite{dvorak2016using, Becker2016}.
%, schmidt2007systematic
 These works neglect precedence constraints and, in terms of scheduling, consider each core or interconnect to be scheduled independently, unlike the co-scheduling of cores and interconnects in this article. The advantage of co-scheduling lies in synchronization between tasks executing on the cores and messages transmitted through an interconnect that results in high efficiency of the system in terms of resource utilization. Steiner~\cite{Steiner} introduces precedence dependencies between activities, while dealing with the problem of scheduling a TTEthernet network. However, Steiner assumes that all activities have identical processing times, which in our case will increase resource utilization significantly.

Some works deal with JC scheduling without any constraints on jitter requirements, which is not realistic in the automotive domain, since there can be jitter-sensitive activities. Puffitsch et al. in~\cite{puffitsch2015off} assume a platform with imperfect time synchronization and propose an exact constraint programming approach. Abdelzaher and Shin in~\cite{abdelzaher1999combined} solve a similar problem by applying both an optimal and a heuristic branch-and-bound method. Furthermore, the authors in~\cite{peng1989static} consider the preemptive version of our problem that makes it impossible to apply their solution to problem considered in this article, since some activities can be scheduled with interruption. 

Jitter requirements are not considered in the problem formulations of~\cite{monnier1998genetic} and~\cite{ramam}, where the authors propose heuristic algorithms to deal with the co-scheduling problem. Finally, in~\cite{di2000scheduling} the authors solve the considered problem with an objective to minimize the jitter of the activities using simulated annealing, while we rather assume jitter-constrained activities with no objective to optimize. Note that these approaches with JC assumption are heuristics and the efficiency of the proposed methods have not been compared to optimal solutions.

There also exist works that schedule both tasks and messages, while assuming ZJ scheduling. Lukasiewycz and Chakraborty~\cite{lukasiewycz12} solve the co-scheduling problem assuming the interconnect to be a FlexRay bus, which results in a different set of constraints. Their approach involves decomposing the initial problem and solving the smaller parts by an ILP approach to manage scalability. Besides, Lukasiewycz et al. in~\cite{Lukasie2012} solve the co-scheduling problem by introducing the preemption into the model formulation. Moreover, Craciunas and Oliver~\cite{craciunas2015combined} consider an Ethernet-based interconnect and solve the problem using both SMT and ILP. However, ZJ scheduling results in a larger number of required cores, as shown in Section~\ref{experiment}. In summary, \emph{this work is different in that it is the first to consider the periodic JC co-scheduling problem with jitter requirements and solves it by a heuristic approach, whose quality is evaluated by comparing with the exact solution for smaller instances}.

%Thus, scheduling in the jitter-constrained manner is a trade-off between memory consumption and the number of required resources that is typically about tens or hundreds of kilobytes of memory versus an additional core or an interconnect.
\begin{comment}
\begin{itemize}
\item There exist two fundamental approaches to solve the problem of periodic scheduling in real-time systems that involves hard real-time requirements. The first one is so called \textit{Event-Triggered Scheduling}, where scheduling is done on-line, triggered by events in the system. This approach is not always suitable for solving scheduling problems in safety-critical systems due to complex predictability of the systems behavior. Works that apply event-triggered strategy for periodic scheduling problems are~\cite{Davis:2011:SHR:1978802.1978814,Baruah1996,Jeffay}. The second approach is \textit{Time-Triggered Scheduling} that we target, where the schedule of the resource allocation is constructed off-line and tasks are scheduled on resources at precise times that are given by a solution (schedule). Examples of works, dealing with periodic scheduling problem and employing time-triggered approach are~\cite{pinwheel,Michelon,Hanzalek2010,Korst1991,BarNoy2004155,ramam}. ~\cite{goswami2012time,zhang2014task,hu2015scheduling}

%Moreover, none of the mentioned work validate their approach on use-cases, inspired by real automotive systems.
\end{itemize}
\end{comment}

%%%%%%%%%%%%%%%%%%%%%%%%%%%%%%%%%%%%%%%%%%%%%%%%%%%%%%%%%%%%%%%
\section{System Model}
\label{description}
This section first introduces the platform and the application models used in this article. Then, the mapping of activities to resources is described, concluded by the problem statement.

\subsection{Platform Model}

The considered platform comprises a set of homogeneous \emph{cores} on a single multi-core Electronic Control Unit (ECU) with a \emph{crossbar switch}, as shown in Figure~\ref{fig:problem}. This is similar to the TriCore architecture~\cite{TriCore}. The crossbar switch provides point-to-point connection between the cores, and input ports act as communication endpoints and can receive only a single message at a time. We assume that tasks on different cores communicate via the crossbar switch that writes variables in local memories of the receiving cores. On the other side, intra-core communication is realized through reading and writing variables that are stored in the local memory of each core. The set of $\numResources$ resources that include $\frac{\numResources}{2}$ cores and $\frac{\numResources}{2}$ crossbar switch input ports is denoted by $\resourceSet = \{ \resource_1,\resource_2,\cdots,\resource_{\numResources} \}$. Moreover, the cores are characterized by their clock frequency and available memory capacity.

Although this work focuses on multi-core systems with crossbar switches, the current formulation is easily extensible to distributed architectures with multiple single-core processing units, connected by a bus, e.g. CAN~\cite{CANBosch}. Furthermore, assuming systems with fully switched networks, e.g. scheduling of time-triggered traffic in TTEternet~\cite{steiner2008ttethernet} leads to a similar scheduling problem. However, scalability of the solution presented below may be problematic in such case due to the increased number of entities to schedule, since each message needs to be scheduled on every network segment. The possible extension of this article in this direction can be found in~\cite{Hanzalek2010}, where scheduling of only communication is done unlike the co-scheduling considered in this article. 

\subsection{Application Model}
The application model is based on characteristics of realistic benchmarks of a specific modern automotive software system, provided in~\cite{WATERSAutomotive2015}. We model the application as a set of periodic \emph{tasks} $\taskSet$ that communicate with each other via a set of \emph{messages} $\messageSet$, transmitted over the crossbar switch. Then $\activitySet = \taskSet \cup \messageSet$, denotes the set of activities, which includes both the incoming messages on the input ports of the crossbar switch and the tasks executed on the cores.
Each activity $\activity_i$ is characterized by the tuple $\{\period_i, \executionTime_i,\jitter_i \}$ representing its period, execution time and jitter requirements, respectively. Its execution may not be preempted by any other activity, since \emph{non-preemptive scheduling} is considered. The release date of each activity equals the start of the corresponding period and the deadline is the end of the next period. This deadline prolongation extends the solution space. The period of a message is set to the period of the task that sends the message. Additionally, execution time of messages on the input ports correspond to the time it takes to write the data to the local memory of the receiving core. Thus, since the local memories are defined by both their bandwidth and latency, execution time for each message $\activity_i \in \messageSet$ is calculated as $\executionTime_i = \frac{sz_i}{bnd} + lat$, where $sz_i$ is the size of the corresponding transmitted variable given by the application model, while $bnd$ is the bandwidth of the memory and $lat$ is its latency given by the platform model. This is similar to latency-rate server concept~\cite{Minaeva201644}. %Stiliadis98, 

%The jitter requirements of messages are set to be equal to their periods, since if a message is a part of a cause-effect chain and the runnable, preceding and following the message have a critical jitter constraint, the message is also forced to have a critical jitter constraint because of the precedence relationships.

\begin{figure}
\centering
\epsfig{file=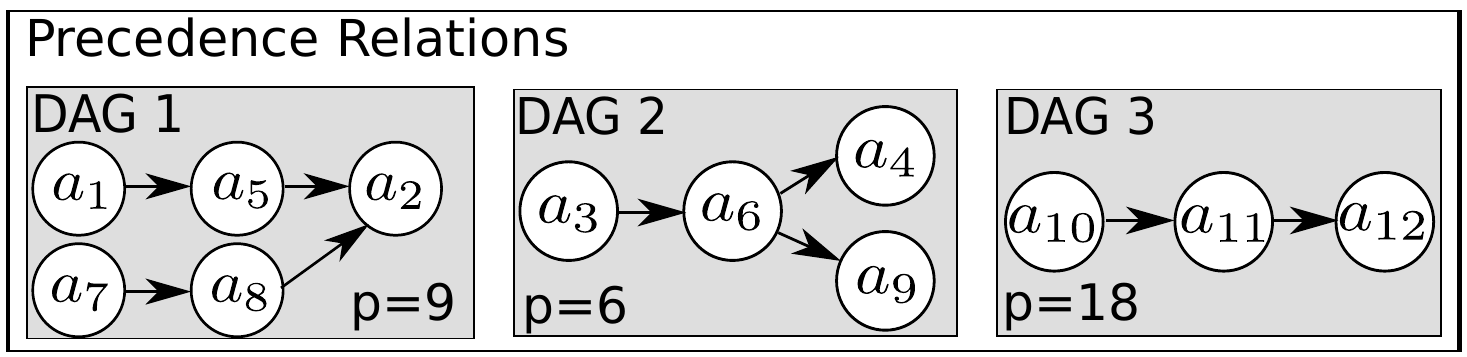,width=1.0\columnwidth}
\caption{An example of the resulting precedence relations, where activities in DAG 1 have period 6, activities in DAG 2 have period 9 and activities in DAG 3 have period 18.}
\label{fig:DAG}
\end{figure}

%\subsubsection{Cause-Effect Chains} 
\emph{Cause-effect chains} are an important part of the model. A cause-effect chain comprises a set of activities that must be executed in a given order within predefined time bounds to guarantee correct behavior of the system. As one activity can be a part of more than one cause-effect chain, the resulting dependency relations are represented by a directed acyclic graph (DAG) that can be very complex in real-life applications~\cite{panic2014runpar}, such as automotive engine control. Similarly to~\cite{craciunas2015combined} and~\cite{hu2015scheduling}, activities of one cause-effect chain are assumed to have the same period. More generalized precedence relations with activities of distinct periods being dependent on each other can be found in e.g.~\cite{forget2011dynamic} . Thus, the resulting graph of precedence relations consists of distinct DAG's for activities with different periods, although not necessarily only one DAG for each unique period. The set of \emph{precedence relations} between the activities is characterized by an adjacency matrix with elements $\adjMatrix = \{\precedenceGraphMatrix_{i,l}\}$ of dimension $\nActivities \times \nActivities$ where $\precedenceGraphMatrix_{i,l} = 1$ if activity $\activity_i$ must finish before activity $\activity_l$ can start. An example of a precedence relation is shown in Figure~\ref{fig:DAG}, which includes the activities from Figure~\ref{fig:example1}. Note that many activities may not have any precedence constraints, since they are not part of any cause-effect chain. For instance, it could be simple logging and monitoring activities.

Lastly, each cause-effect chain has an \emph{end-to-end deadline} constraint, i.e. the maximum time that can lapse from the start of the first activity till the end of the execution of the last activity in each chain equal to two corresponding periods. However, as the first activity in each chain can be scheduled at the beginning of the period at the earliest and the last activity of the chain at the end of the next period at the latest, the end-to-end latency constraint is automatically satisfied due to release and deadline constraints of the activities. Therefore, end-to-end latency constraints do not add further complexity to the model.

\subsection{Mapping of Tasks to Cores}
\label{mapping}
The mapping $\mapping:\activitySet \rightarrow \resourceSet$, $\mapping = \{\mapping_1,\cdots, \mapping_{\nActivities}\}$ of tasks to cores and messages to the memories is assumed to be given by the system designer, which reflects the current situation in the automotive domain, for e.g. engine control. Note that for the previously discussed extension to fully switched network systems, both mapping and routing (i.e. define path through the network for each message) that respect some locality constraints are necessary. Since mapping influences routing and therefore message scheduling, for such systems it is advantageous to solve the three steps at once, as it is done e.g. in~\cite{tuamacs2015design}.
  
To get a mapping for the problem instances used to validate the approaches in this article, a simple ILP model for mapping tasks to cores is formulated the following way. The variables $\mapVariable_{i,j} \in \{0,1\}$ indicate whether task $i = 1,\cdots,|\taskSet|$ is mapped to resource $j = 1,\cdots,\frac{\numResources}{2}$ ($\mapVariable_{i,j}$ = 1) or not ($\mapVariable_{i,j}$ = 0). Note that we consider only Cores as resources and tasks on cores as activities while mapping. The mapping tries to balance the load, which is formulated as a sum of absolute values of utilization differences on two consecutive resources in Equation~\eqref{eq:mappingCrit}. Since the absolute operator is not linear, it needs to be linearized by introducing the load variables $\loadBalance_{j} \in \rationalNumberSet$ in Equations~\eqref{map1}~and~\eqref{map2}. 
\vspace{-0.1cm}
\begin{equation}
\label{eq:mappingCrit}
\textit{Minimize:} \; \sum_{j \in 1,..,\frac{\numResources}{2}} \loadBalance_{j}
\end{equation} 
\emph{subject to}:
\begin{equation}
\label{map1}
\loadBalance_{j} \geq \sum_{i = 1,\cdots,|\taskSet|} \frac{\executionTime_i}{\period_i} \cdot \mapVariable_{i,j} - \sum_{i = 1,\cdots,|\taskSet|} \frac{\executionTime_i}{\period_i} \cdot \mapVariable_{i,j+1}, \; j = 1,\cdots,\frac{\numResources}{2}
\end{equation} 
\vspace{-0.2cm}
\begin{equation}
\label{map2}
\loadBalance_{j} \geq \sum_{i = 1,\cdots,|\taskSet|} \frac{\executionTime_i}{\period_i} \cdot \mapVariable_{i,j + 1} -  \sum_{i = 1,\cdots,|\taskSet|} \frac{\executionTime_i}{\period_i} \cdot \mapVariable_{i,j}, \; j = 1,\cdots,\frac{\numResources}{2}.
\end{equation}
Moreover, each task must be mapped to a single resource, as is stated in Equation~\eqref{map3}. 
\begin{equation}
\label{map3}
\sum_{j = 1,\cdots,\frac{\numResources}{2}} \mapVariable_{i,j} = 1, \; i = 1,\cdots,|\taskSet|. \\
\end{equation}
%\vspace{-0.2cm}
%\begin{equation}
%\label{map4}
%\sum_{i = 1,\cdots,|\taskSet|} \frac{\executionTime_i}{\period_i} \cdot \mapVariable_{i,j} \leq \utilizationBound, \; j = 1,\cdots,\numResources. \\
%\end{equation}
%
Note that this mapping is not considered a contribution of this article, but only a necessary step to provide a starting point for the experiments, since the benchmark generator does not provide the mapping. 
 
\subsection{Problem Statement}
\label{statement}

Given the above model, the goal is to find a schedule with a hyper-period $\hyperPeriod = lcm(\period_1, \period_2, \cdots, \period_{\nActivities})$ with $lcm$ being the least common multiple function, where the schedule is defined by start times $\scheduleOrderStart_{i}^{j} \in \naturalNumberSet$ of each activity $\activity_i \in \activitySet$ in each period $j = 1,2,\cdots,\numActivityOccurrence_i$, where $\numActivityOccurrence_i = \frac{\hyperPeriod}{\period_i}$. The schedule must satisfy the periodic nature of the activities, the precedence relations and the jitter constraints. The considered scheduling problem can be categorized as \emph{multi-periodic non-preemptive scheduling of activities with precedence and jitter constraints on dedicated resources}. 

The formal definition of a zero-jitter schedule is the following:
\begin{definition}[Zero-jitter (ZJ) schedule]
\label{def:ZJS}
The schedule is a ZJ schedule if and only if for each activity $\activity_i$ Equation~\eqref{eq:ZJ} is valid, i.e. the difference between the start times $\scheduleOrderStart_i^{j}$ and $\scheduleOrderStart_i^{j+1}$ in each pair of consecutive periods $j$ and $j+1$ over the hyper-period is the same.
\begin{equation}
\label{eq:ZJ}
\scheduleOrderStart_{i}^{j+1} - \scheduleOrderStart_i^{j} = \period_i, \; j = 1,2,\cdots,\numActivityOccurrence_i-1.
\end{equation}
\end{definition}
Zero-jitter scheduling deals exclusively with ZJ schedules. If for some activity and some periods $j$ and $j+1$ Equation~\eqref{eq:ZJ} does not hold in the resulting schedule, we call it \emph{jitter-constrained (JC) schedule}.

The scheduling problem, where a set of periodic activities are scheduled on one resource is proven to be NP-hard in~\cite{cai1996nonpreemptive} by transforming from the 3-Partition problem. Thus, the problem considered (both ZJ and JC) here is also NP-hard, since it is a generalization of the aforementioned NP-hard problem.

%%%%%%%%%%%%%%%%%%%%%%%%%%%%%%%%%%%%%%%%%%%%%%%%%%%%%%%%%%%%%
\section{Exact Models}
\label{ILP}
Due to significantly different timing behavior of the models on problem instances with varying complexity (see Section~\ref{experiment}), both SMT and ILP models are formulated in this article. Moreover, the NP-hardness of the considered problem justifies using these approaches, since no polynomial algorithm exists to optimally solve the problem unless P=NP. This section first presents a minimal SMT formulation to solve the problem optimally, then continues with a linearization of the SMT model to get an ILP model. It concludes by providing improvements to both models that exploit problem-specific knowledge, reducing the complexity of the formulation and thus the computation time.

\subsection{SMT Model} 
The SMT problem formulation is based on the set of variables $\scheduleOrderStart_i^j \in \{1,2,\cdots,\hyperPeriod\}$, indicating a start time of job $j$ of activity $\activity_i$. Following the problem statement in Section~\ref{statement}, we deal with a decision problem with no criterion to optimize. The solution space is defined by five sets of constraints. The first set of constraints is called \emph{release date and deadline constraints} and it requires each activity to be executed in a given time interval of two periods, as stated in Equation~\eqref{constrPD1}. 
\begin{align}
\label{constrPD1}
(j - 1) \cdot \period_i \leq \scheduleOrderStart_{i}^{j} \leq (j + 1) \cdot \period_i - \executionTime_i,  \\ \activity_i \in \activitySet, \ j = 1,\cdots, \numActivityOccurrence_i. \nonumber
\end{align}
The second set, Constraint~\eqref{RC}, ensures that for each pair of activities $\activity_{i}$ and $\activity_{l}$ mapped to the same resource ($\mapping_{i} = \mapping_{l}$), it holds that either $\activity_{i}^j$ is executed before $\activity_{l}^k$ or vice-versa. These constraints are called \emph{resource constraints}. Note that due to the extended deadline in Constraint~\eqref{constrPD1}, the resource constraints must be added also for jobs in the first period with jobs of the last period, since they can collide. 
\begin{align}
\label{RC}
\begin{split}
\scheduleOrderStart_i^j +  \executionTime_{i} \leq \scheduleOrderStart_l^k \lor \scheduleOrderStart_l^k +  \executionTime_{l} \leq \scheduleOrderStart_i^j , \qquad \qquad \\
\scheduleOrderStart_i^1 +  \executionTime_{i} + \hyperPeriod \leq \scheduleOrderStart_l^{\numActivityOccurrence_l} \lor \scheduleOrderStart_l^{\numActivityOccurrence_l} +  \executionTime_{l} \leq \scheduleOrderStart_i^1 + \hyperPeriod, \qquad \qquad \\
\activity_{i}, \ \activity_{l} \in \activitySet: \mapping_{i} = \mapping_{l}, \ j = 1,...,\numActivityOccurrence_i, \ k = 1,...,\numActivityOccurrence_l. 
\end{split}
\end{align}
For the ZJ case, it is enough to formulate Constraints~\eqref{RC} for each pair of activities only for jobs in the least common multiple of their periods, i.e. $j = 1,\cdots,\frac{lcm(\period_i, \period_j)}{\period_i}$ and $k = 1,\cdots,\frac{lcm(\period_i, \period_j)}{\period_l}$. Moreover, the problem for ZJ scheduling is formulated using $\nActivities$ variables. One variable $\scheduleOrderStart_i^1$ is defined for the first job of each activity and other jobs are simply rewritten as $\scheduleOrderStart_{i}^{j} = \scheduleOrderStart_{i}^{1} + \period_i \cdot (j-1)$.

The next set of constraints is introduced to prevent situations, when two consecutive jobs of one activity collide. Thus, Constraint~\eqref{constrPD5} introduces precedence constraints between each pair of consecutive jobs of each activity, considering also the last and the first job. 
\begin{align}
\label{constrPD5}
\begin{split}
\scheduleOrderStart_{i}^{j} + \executionTime_i \leq \scheduleOrderStart_{i}^{j + 1},   \qquad  \\ 
\scheduleOrderStart_{i}^{\numActivityOccurrence_i} + \executionTime_i \leq \scheduleOrderStart_{i}^{0} + \hyperPeriod,   \qquad  \\   
  \activity_i \in \activitySet, \ j = 1,\cdots, \numActivityOccurrence_i - 1.  
  \end{split}
\end{align} 
%
%Note that this constraint does not break optimality, since if job $j+1$ and its successors can be scheduled, job $j$ with its successors can be scheduled at their place due to the almost identical requirements of each pair of jobs. The only difference in the requirements of the consecutive jobs of one activity are the release date and deadline constraints, which are either trivially satisfied or the considered instance is infeasible. Furthermore, this constraint eliminates symmetries in the solution space, resulting in reduced computation time. \todo{is this understandable description and how to improve it? Maybe it's worthwhile to include the formal proof?}

Next, due to the existence of cause-effect chains, \emph{precedence constraints} that are based on the previously mentioned $\adjMatrix$ matrix are formulated in Equation~\eqref{PRC}. 
\begin{align}
\label{PRC}
\scheduleOrderStart_{i}^{j} + \executionTime_i \leq \scheduleOrderStart_{l}^{j},   \qquad \qquad \\  \activity_i, \ \activity_l \in \activitySet: \precedenceGraphMatrix_{i,l} = 1 , \ j = 1,\cdots, \numActivityOccurrence_i.  \nonumber
\end{align}
The \emph{jitter constraints} can be formulated either in terms of \emph{relative jitter}, where we bound only the difference in start times of jobs in consecutive periods or in terms of \emph{absolute jitter}, bounding the start time difference of any two jobs of an activity. Experiments have shown that defining jitter as absolute or relative does not significantly influence the resulting efficiency. The difference in terms of maximal achievable utilization is less than 1\% on average with relative jitter showing higher utilization. Therefore, further in the paper we use the relative definition of jitter. Note that the results for absolute jitter formulation do not differ significantly from the results presented in this article. The formulation of relative jitter is given in Equation~\eqref{constrPD3}, where the first constraint deals with jitter requirements of jobs inside one hyper-period and the second one deals with jobs crossing a border between two hyper-periods. 
\begin{align}
\begin{split}
\label{constrPD3}
|\scheduleOrderStart_{i}^{j} - (\scheduleOrderStart_{i}^{j - 1} + \period_i) | \leq \jitter_i, \\
|\scheduleOrderStart_{i}^{1} + \hyperPeriod - \period_i - \scheduleOrderStart_{i}^{\numActivityOccurrence_i} | \leq \jitter_i \\
j = 2,\cdots, \numActivityOccurrence_i: \ j > k, \ \activity_i \in \activitySet. 
\end{split}
\end{align}

\subsection{ILP Model}
\label{ILP_model}
The formulation of the ILP model is very similar to the SMT model described above. The main difference in formulation is caused by the requirement of linear constraints for the ILP model. Thus, since Equations~\eqref{constrPD1},~\eqref{constrPD5} and~\eqref{PRC} are already linear, they can be directly used in the ILP model. However, resource Constraints~\eqref{RC} are non-linear and to linearize them, we introduce new set of decision variables that reflect the relative order of each two jobs of different activities: 
\begin{equation*}
\scheduleOrder_{i,l}^{j,k} =
  \begin{cases}
    1, & \text{if $\activity_{i}^j$ starts before $\activity_{l}^k$};\\
    0, & \text{otherwise}.
  \end{cases}
\end{equation*}  
Therefore, resource constraints are formulated by Equation~\eqref{limitedLCMRO1}, which ensures that either $\activity_{i}^j$ is executed before $\activity_{l}^k$ (the first equation holds and $\scheduleOrder_{i,l}^{j,k} = 1$) or vice-versa (the second equation holds and $\scheduleOrder_{i,l}^{j,k} = 0$). However, exactly one of these equations must always hold due to binary nature of $\scheduleOrder_{i,l}^{j,k}$, which prevents the situation where two activities execute simultaneously on the same resource. Note that we use $2 \cdot \hyperPeriod$ in the right part of the Constraints, since the maximum difference between two jobs of distinct activities can be maximally $2 \cdot \hyperPeriod$ due to release date and deadline constraints.
\begin{align}
\label{limitedLCMRO1}
\begin{split}
\scheduleOrderStart_i^j +  \executionTime_{i} \leq \scheduleOrderStart_l^k + 2 \cdot \hyperPeriod \cdot (1 - \scheduleOrder_{i,l}^{j,k}), \qquad \\
\scheduleOrderStart_l^k +  \executionTime_{l} \leq \scheduleOrderStart_i^j + 2 \cdot \hyperPeriod \cdot \scheduleOrder_{i,l}^{j,k}, \qquad  \\
\activity_{i}, \ \activity_{l} \in \activitySet, j = 1,...,\numActivityOccurrence_i, \ k = 1,...,\numActivityOccurrence_l. 
\end{split}
\end{align}
Furthermore, to formulate the jitter constraints~\eqref{constrPD3} in a linear form, the absolute value operator needs to be eliminated. As a result, Equation~\eqref{limitedLCMRO7} introduces four sets of constraints, two for the jobs inside one hyper-period and two for the jobs on the border.
\begin{align} 
\label{limitedLCMRO7}
\begin{split}
\scheduleOrderStart_{i}^{j} - (\scheduleOrderStart_{i}^{k} + (j - k) \cdot \period_i) \leq \jitter_i, \\ \scheduleOrderStart_{i}^{j+1} - (\scheduleOrderStart_{i}^{j} + (j - k) \cdot \period_i)  \geq  - \jitter_i   \\  
(\scheduleOrderStart_{i}^{1} + \hyperPeriod - \period_i) - \scheduleOrderStart_{i}^{\numActivityOccurrence_i} \leq  \jitter_i \\
(\scheduleOrderStart_{i}^{1} + \hyperPeriod - \period_i) - \scheduleOrderStart_{i}^{\numActivityOccurrence_i} \geq  -\jitter_i \\
j, k = 1,\cdots, \numActivityOccurrence_i: \ j > k, \ \activity_i \in \activitySet. 
\end{split}
\end{align}
Unlike the time-indexed ILP formulation~\cite{kone2011event}, where each variable $y_{i,j}$ indicates that the activity $i$ is scheduled at time $j$ (having $\hyperPeriod \cdot \nActivities$ variables), the approach used here can solve problems with large hyper-periods when there are fewer jobs with longer execution time. Hence, it utilizes only $\nJobs + \frac{\nJobs \cdot (\nJobs - 1)}{2}$ with $\nJobs = \sum_{i=1}^{\nActivities}{\numActivityOccurrence_i}$ variables, which is a fraction of the variables that the time-indexed formulation requires for this problem.

%%%%%%%%%%%%%%%%%%%%%%%%%%%%%%%%%%%%%%%%%%%%%%%%%%%%%%%%%%%%
\subsection{Computation Time Improvements}
\label{optimizations}
While the basic formulations of the SMT and ILP models were presented previously, four computation time improvements for the models are introduced here in order to reduce the complexity of the formulation and computation time of the solver. Note that the improvements do not break the optimality of the solution. 

The first improvement \emph{removes redundant resource constraints}. Due to the release date and deadline constraints~\eqref{constrPD1}, it is known that $\scheduleOrderStart_{i}^{j} \in [(j-1)\cdot \period_i, (j + 1)\cdot \period_i - \executionTime_i]$ and $\scheduleOrderStart_{l}^{k} \in [(k-1)\cdot \period_l, (k + 1) \cdot \period_l - \executionTime_l]$. Therefore, it is necessary to include resource constraints only if the intervals overlap. 
%Finally, if two activities have a precedence relation, i.e. they are a part of the same DAG component, it is not necessary to have resource constraints, since they are guaranteed by precedence constraints. 
This improvement results in more than 20\% of the resource constraints being eliminated, reducing the computation time significantly since the number of resource constraints grows quadratically with the number of activities mapped to a given resource.

Instead of setting the release date and deadline Constraint~\eqref{constrPD1}, the second improvement provides this information directly to the solver. Thus, each constraint is substituted by setting the lower bound of $\scheduleOrderStart_{i}^{j}$ on $(j - 1) \cdot \period_i$ and the upper bound on $(j + 1) \cdot \period_i - \executionTime_{i}$. Hence, instead of assuming the variables $\scheduleOrderStart_{i}^{j}$ in interval $[1,\cdots, \hyperPeriod]$ and pruning the solution space by the periodicity constraints, the solver starts with tighter bounds for each variable. This significantly cuts down the search space, thereby reducing computation time. Due to the different solver abilities for SMT and ILP, this optimization is only applicable to the ILP model.

We can further \emph{refine the lower and upper bounds} of the variables by exploiting the knowledge about precedence constraints, which is the third improvement. For each activity the length of the longest critical path of the preceding and succeeding activities that must be executed \emph{before} and \emph{after} the given activity, $\criticalTimeBefore$ and $\criticalTimeAfter$ respectively, are computed. First, the values of $\lowerBound{\criticalTimeBefore_i}$ and $\lowerBound{\criticalTimeAfter_i}$ are obtained by adding up the execution times of the activities in the longest chain of successors and predecessors of the activity $\activity_i$, respectively, as proposed by~\cite{chetto1990dynamic}. For the example in Figure~\ref{fig:DAG}, assuming the execution times of all activities are equal to 1,  $\lowerBound{t^b_1} = 0$, $\lowerBound{t^a_1} = 2$, $\lowerBound{t^b_6} = 1$, $\lowerBound{t^a_6} = 1$, $\lowerBound{t^b_2} = 2$, $\lowerBound{t^a_2} = 0$. Additionally, the bounds can be improved by computing the sum of execution times of all the predecessors, mapped to the same resource, i.e. 
$$\criticalTimeBefore_i = \max(\sum_{l: \ l \in Pred_i, \ \mapping_l = \mapping_i} \executionTime_l, \lowerBound{t^b_i})$$ \vspace{-0.2cm}
$$\criticalTimeAfter_i = \max(\sum_{l: \ l \in Succ_i, \ \mapping_l = \mapping_i} \executionTime_l, \lowerBound{t^a_i}),$$ 
where $Pred_i$ and $Succ_i$ denote the set of all predecessors and all successors of activity $\activity_i$, respectively.

For the example in Figure~\ref{fig:DAG} and a single core, the resulting values are the following:  $\criticalTimeBefore_1 = 0$, $\criticalTimeAfter_1 = 2$, $\criticalTimeBefore_6 = 1$, $\criticalTimeAfter_6 = 2$, $\criticalTimeBefore_2 = 4$, $\criticalTimeAfter_2 = 0$. Hence, the lower bound of $\scheduleOrderStart_{i}^{j}$ can be refined by adding $\criticalTimeBefore_i$ and the upper bound can be tightened by subtracting $\criticalTimeAfter_i$, i.e. $ \scheduleOrderStart_{i}^{j} \in [(j-1)\cdot \period_i + \criticalTimeBefore_i, (j + 1) \cdot \period_i - \executionTime_i - \criticalTimeAfter_i]$. This can also be used in the first improvement, eliminating even more resource constraints.

%, since the predecessors (successors) always have to be executed before (after) an activity.

The fourth and final improvement \emph{removes jitter constraints~\eqref{limitedLCMRO7} for activities with no freedom to be scheduled with larger jitter than required}. For instance, for jobs of $\activity_2$ from Figure~\ref{fig:DAG} with $\executionTime_2 = 1$, $\criticalTimeBefore_2 = 4$, $\criticalTimeAfter_2 = 0$ and $\period_2 = 9$, there are only 14 instants $t$, where it can be scheduled, i.e. $t \in \{4,\cdots, 17\}$. If $\jitter_2 \geq 13$, the jitter constraint can be omitted since the activity can be scheduled only at 14 instants due to the third improvement and it is not possible to have jitter bigger than 13 time units and still respect the periodicity of the activity. We denote by $\lengthOfSchedInterval_i$ the worst-case slack of the activity, i.e. the lower bound on the number of time instants where activity $\activity_i$ can be scheduled and we compute it according to Equation~\eqref{opt1}. Hence, the jitter constraints are only kept in the model if Inequality~\eqref{opt2} holds, i.e. the activity has space to be scheduled with larger jitter than required. We refer to an activity satisfying Equation~\eqref{opt2} \emph{jitter-critical}. Otherwise, it is a \emph{non-jitter-critical} activity.
\begin{gather}
\label{opt1}
\lengthOfSchedInterval_i = \period_i - (\criticalTimeBefore + \criticalTimeAfter + \executionTime_i)\\
\label{opt2}
\jitter_i \leq \lengthOfSchedInterval_i - 2, \ \activity_i \in \activitySet.
\end{gather}
Experimental results have shown that even on smaller problem instances with 40-55 activities, the proposed improvements reduce computation time by up to 30 times for ILP model and 12 times for SMT model. Moreover, the first and the third improvements result in the most significant reduction of the computation time. However, when experimentally comparing these two improvements, we see that the behavior is rather dependent on the problem instance characteristics, as both the first and the third improvements can be the most effective on different problem instances.

%%%%%%%%%%%%%%%%%%%%%%%%%%%%%%%%%%%%%%%%%%%%%%%%%%%%%%%%%%%%%%
\section{Heuristic Algorithm}
\label{heuristic}
Although the proposed optimal models solve the problem optimally, this section introduces a heuristic approach to solve the problem in reasonable time for \emph{larger instances}, possibly sacrificing the optimality of the solution within acceptable limits. 

\subsection{Overview} 

\begin{figure}
\centering
\epsfig{file=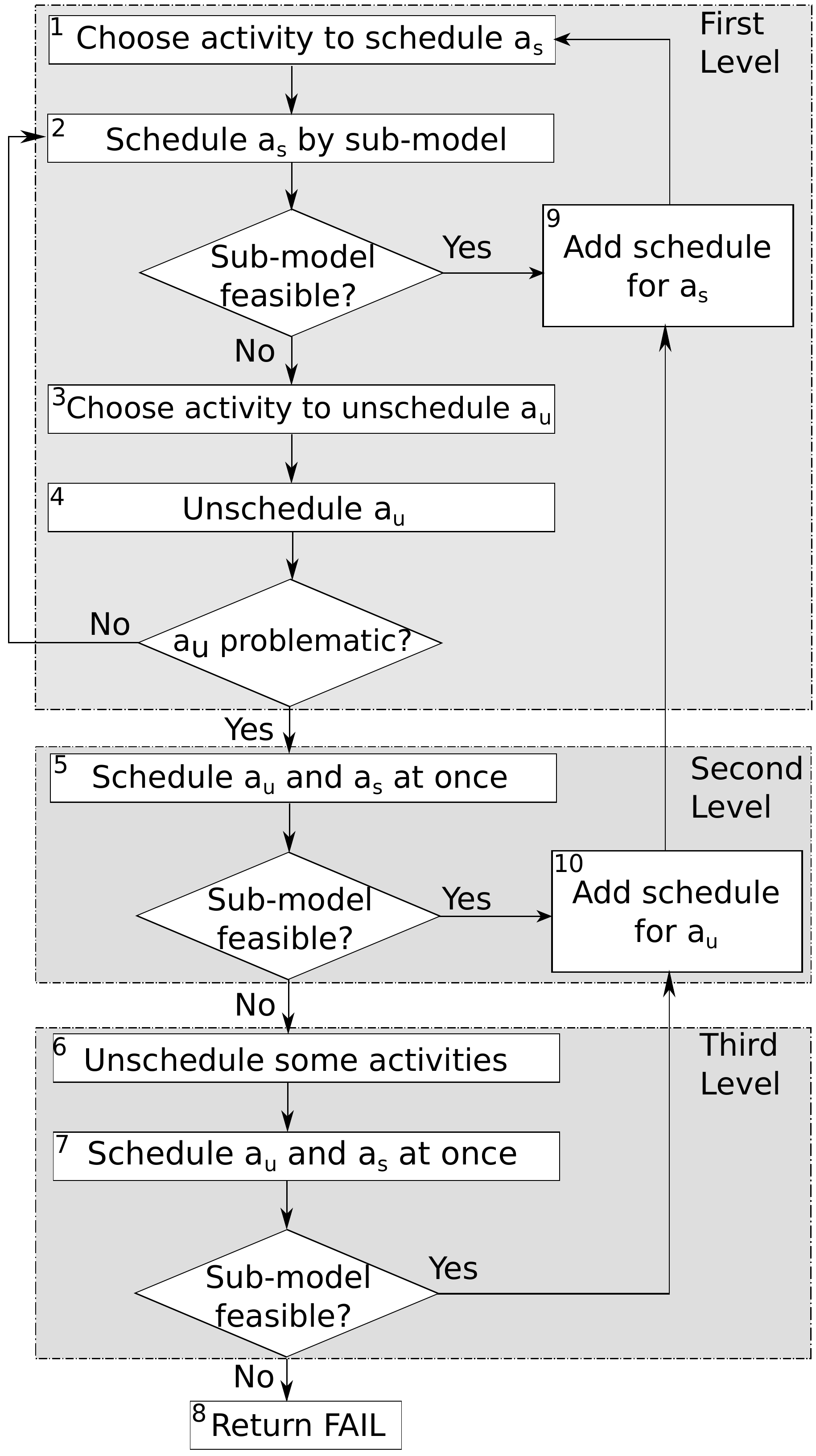,width=0.9\columnwidth}
\caption{Outline of 3-Level Scheduling heuristic.}
\label{fig:heuristic_description}
\end{figure}

The proposed heuristic algorithm, called \emph{3-Level Scheduling (3-LS)} heuristic, creates the schedule constructively. It assigns the start time to every job of an activity in a given HP. Moreover, it implements 3 levels of scheduling, as shown in Figure~\ref{fig:heuristic_description}. The first level inserts activity by activity into the schedule, while removing some of the previously scheduled activities, $\activityToUnschedule$, if the currently scheduled activity $\activityToSchedule$ cannot be scheduled. However, in case the activity $\activityToUnschedule$ to be removed had problems being scheduled in previous iterations, the algorithm goes to the second level, where two activities that were problematic to schedule, $\activityToSchedule$ and $\activityToUnschedule$ are \emph{scheduled simultaneously}. By scheduling these two activities together, we try to avoid problems with a sensitive activity further in the scheduling process. Simultaneous scheduling of two activities means that two sets of start times $\scheduleOrderStart_c$ and $\scheduleOrderStart_u$ are decided for activities $\activityToSchedule$ and $\activityToUnschedule$ concurrently. The third scheduling level is initiated when even co-scheduling two activities $\activityToSchedule$ and $\activityToUnschedule$ simultaneously does not work. Then, the third level starts by removing all activities except the ones that were already scheduled by this level previously and its predecessors. Next, it co-schedules the two problematic activities again. Note that although there may be more than two problematic activities, the heuristic always considers maximally two at once.

Having three levels of scheduling provides a good balance between solution quality and computation time, since the effort to schedule problematic activities is reasonable to not prolong the computation time of the approach and to get good quality solutions. Experimental results show that 94\% of the time is spent in the first scheduling level, where the fastest scheduling takes place. However, in case the first level does not work, the heuristic algorithm continues with the more time demanding second scheduling level and according to the experimental results it spends 3\% of time in this level. The final 3\% of the total computation time is spent in the third scheduling level that prolongs the computation time the most since it unschedules nearly all the activities scheduled before. Thus, three levels of scheduling is a key feature to make the heuristic algorithm cost efficient and yet still able to find a solution most of the time. As seen experimentally in Section~\ref{experiment}, it suffices to find a good solutions for large instances within minutes.

Note that the advantage of scheduling all jobs of one activity at a time compare to scheduling by individual jobs lies in the significantly reduced number of entities we need to schedule. Hence, unlike the exact model that focus on scheduling jobs for all of the activities at a time, the 3-LS heuristic approach decomposes the problem to smaller sub-problems for one activity. This implies that the 3-LS heuristic is not optimal and is also a reason why it takes significantly less time to solve the problem.
\subsection{Sub-model} 
The schedule for a single activity or two activities at the same time, respecting the previously scheduled activities is found by a so called \emph{sub-model}. The sub-model for one activity $\activity_i$ that is non-jitter-critical (i.e. $\activity_i$ for which Inequality~\eqref{opt2} does not hold) is formulated as follows. The minimization criterion is the sum of the start times of all $\activity_i$ jobs (Equation~\eqref{eq:heurCrit}). Note that the activity index $i$ is always fixed in the sub-model since it only schedules a single activity at a time.
\begin{equation}
\label{eq:heurCrit}
\textit{Minimize:} \; \sum_{j \in 1..\numActivityOccurrence_i} \scheduleOrderStart_i^{j}
\end{equation} 
The reasons for scheduling activities as soon as possible are twofold. Firstly, it is done for dependent activities to extend the time interval in which the successors of the activity can be scheduled, thereby increasing the chances for this DAG component to be scheduled. Secondly, scheduling at the earliest instant helps to reduce the fragmentation of the schedule, i.e. how much free space in the schedule is left between any two consecutively scheduled jobs, resulting in better schedulability in case the periods are almost harmonic, i.e. being multiples of each other, which is common in the automotive domain~\cite{WATERSAutomotive2015}. 

The start time of each job $j$ can take values from the set $D_i^j$ (Equation~\eqref{eq:SMConstr}), which is the union of intervals, i.e. 
\begin{align}
\begin{split}
\label{eq:SMConstr5}
D_i^{j} = \{[l^{i,j}_1,r^{j}_1] \cup [l^{i,j}_2,r^{i,j}_2] \cup \cdots \cup [l^{i,j}_w,r^{i,j}_w] \}, \\
r^{i,j}_o < l^{i,j}_{o+1}, \ l^{i,j}_o \leq r^{i,j}_o, \ j = 1,\cdots,\numActivityOccurrence_i, \ o = 1,\cdots,w - 1
\end{split}
\end{align}
and $\{l^{i,j}_1,r^{i,j}_1,\cdots,l^{i,j}_w,r^{i,j}_w\}~\in~\integerNumberSet^{2 \cdot w}$, where $w$ is the number of intervals in $D_i^{j}$ and $l^{i,j}_o,r^{i,j}_o$ are the start and end of the corresponding interval $o$. This set of candidate start times is obtained by applying periodicity constraints~\eqref{constrPD1} and precedence constraints~\eqref{PRC} to already scheduled activities and changing the resulting intervals so that the activity can be executed fully with respect to its execution time. Note that since we insert only activities whose predecessors are already scheduled, all constraints are satisfied if the start time of the job $\scheduleOrderStart_i^{j}$ belongs to $D_i^{j}$.

For the example in Figure~\ref{fig:DAG} with all the execution times equal to 1, with a single core, and with no activities scheduled%(i.e. $\ScheduledSet = \emptyset$)
, $D_2^{1} =  \{[0 \cdot \period_2 + \criticalTimeBefore_2; 2 \cdot \period_2 - \criticalTimeAfter_2 - \executionTime_2 - 1]\} =  \{[0 + 4; 18 - 0 - 1- 1]\} =  \{[4; 16]\}$, which is basically the application of the third improvement from Section~\ref{ILP}. Now, suppose in the previous iterations $\activity_8^1$ is scheduled at time 4 and $\activity_{10}^1$ is scheduled at time 6. Then, the resulting $D_2^{1} =   \{[5; 5] \cup [7,16]\}$, since $\activity_2^1$ must be scheduled after $\activity_8^1$ and it cannot collide with any other activity on the core.
\begin{equation}
\label{eq:SMConstr}
\scheduleOrderStart_i^{j} \in D_i^{j}, \; j = 1,2,\cdots,\numActivityOccurrence_i.
\end{equation} 
 Furthermore, similarly to the ILP model in Section~\ref{ILP}, the precedence constraints for consecutive jobs of the same activity must also be added.
\begin{align}
\begin{split}
\label{eq:SMConstr2}
\scheduleOrderStart_i^{j} + \executionTime_i \leq \scheduleOrderStart_i^{j + 1},\\
\scheduleOrderStart_i^{\numActivityOccurrence_i} + \executionTime_i \leq \scheduleOrderStart_i^{0} + \hyperPeriod,\\ \; j = 1,2,\cdots,\numActivityOccurrence_i - 1
\end{split}
\end{align} 
The pseudocode of the sub-model is presented in Algorithm~\ref{alg:SM}. As an input it takes the first activity to schedule $\activity_1$, and the optional second activity to schedule $\activity_2$ together with their requirements, and the set of intervals $D$. If $\activity_2$ is set to an empty object, the sub-model must schedule only activity $\activity_1$. In case $\activity_1$ is non-jitter-critical, this scheduling problem can be trivially solved by assigning $\scheduleOrderStart_{i}^j = l^{i,j}_1$ and checking that it does not collide with the job in the previous period. If it does, we schedule this job at the finish time of the previous job if possible, otherwise at the end of the resource interval it belongs to. If the start time is more than the refined deadline of this job from Section~\ref{optimizations}, the activity cannot be scheduled.

It is clear that this rule will always result in a solution, minimizing~\eqref{eq:heurCrit} if one exists. Moreover, if for some job $\scheduleOrderStart_{i}^j$, the interval $D_{i}^j$ is empty, then there is no feasible assignment of this activity to time with the current set of already scheduled activities. 
\begin{algorithm}[ht]
\caption{Sub-model used by 3-LS heuristic}
\label{alg:SM}
\KwIn{$\activity_1$, $\activity_2$, $D$}
\eIf{$\activity_2$ = NULL}{  
		\eIf{$\activity_1$ is non-jitter-critical}{
			$\scheduleHeur = \min_{x \in D_i} x$ : \emph{Constraint~\eqref{eq:SMConstr2} holds}\;
		}{
			$\scheduleHeur = ILP(\activity_1,D)$\;
		}
	}{
		$\scheduleHeur = ILP(\activity_1, \activity_2,D)$\;
	}

\KwOut{$\scheduleHeur$}
\end{algorithm}
On the other hand, when $\activity_1$ is jitter-critical, the sub-model is enriched by the set of jitter constraints~\eqref{limitedLCMRO7} and the strategy to solve it has to be more sophisticated. The sub-model in this case is solved as an ILP model, which has significantly shorter computation times on easier problem instances in comparison to SMT, as shown experimentally in Section~\ref{experiment}. This is important for larger problem instances where sub-model is launched thousands of times, since the heuristic decomposes a large problem to many small problems by scheduling jobs activity by activity. Although this problem seems to be NP-hard in the general case because of the non-convex search space, the computation time of the sub-model is still reasonable due to the relatively small number of jobs of one activity (up to 1000) and the absence of resource constraints.

We formulate Constraint~\eqref{eq:SMConstr} as an ILP in the following way. First, we set $l^{j}_1 \leq \scheduleOrderStart_i^{j} \leq r^{j}_w$ defined earlier in this section, and for each $r^{i,j}_t$ and $l^{i,j}_{t+1}$ two new Constraints~\eqref{heur1} and a variable $y_{i,j,t} \in \{0,1\}$ are introduced, which handles the ``$\lor$'' relation of the two constraints similarly to the variable $\scheduleOrder_{i,l}^{j,k}$ from the ILP model in Section~\ref{ILP}. 
\begin{align}
\begin{split}
\label{heur1}
\scheduleOrderStart_i^{j} + (1 - y_{i,j,t}) \cdot \hyperPeriod \geq l^{i,j}_{t+1}\\
\scheduleOrderStart_i^{j} \leq r^{i,j}_t + y_{i,j,t} \cdot \hyperPeriod  
\end{split}
\end{align}
Finally, when the sub-model is used to schedule two activities at once, i.e. $\activity_2$ is not an empty object, Criterion~\eqref{eq:heurCrit} is changed to contain both activities, and the resource constraints~\eqref{limitedLCMRO1} for $\activity_1$ and $\activity_2$ are added. The resulting problem is also solved as an ILP model, but similarly to the previous case takes rather short time to compute due to small size of the problem. Note that the 3-LS heuristic also utilizes the proposed computation time improvements for the ILP model from Section~\ref{ILP} and always first checks non-emptiness of $D_{j}$ for each job $j$ before creating and running the ILP model.

%%%%%%%%%%%%%%%%%%%%%%%%%%%%%%%%%%%%%%%%%%%%%%%%%%%%%%%%%%%%%%
\subsection{Algorithm}

The proposed 3-LS heuristic is presented in Algorithm~\ref{alg:Heuristic}. The inputs are the \emph{set of activities} $\activitySet$, the \emph{priority rule} $Pr$ that states the order in which the activities are to be scheduled and the \emph{rule to choose the activity to unschedule} $Un$ if some activity is not schedulable with the current set of previously scheduled activities. The algorithm begins by initializing the interval set $D$ for each $\activity_i^j$ as $D_i^j =  \{[(j-1) \cdot \period_i + \criticalTimeBefore_i; j \cdot \period_i - \criticalTimeAfter_i - \executionTime_i - 1]\}$ (line 2). Then it sorts the activities according to the priority rule $Pr$ (line~3), described in detail Section~\ref{rules}. The rule always states that higher priority must be assigned to a predecessor over a successor, so that no activity is scheduled before its predecessors. Note that the first part of the first level of scheduling is similar to the list scheduling approach~\cite{yang1993list}.

In each iteration, the activity with the highest priority $\activityToSchedule$ in the \emph{priority queue of activities to be scheduled} $\queueToSchedule$, is chosen and scheduled by the sub-model (line~7). If a feasible solution $\scheduleHeur$ is found, the interval set $D$ is updated so that all precedence and resource constraints are satisfied. Firstly, for each $\activity_l$ that is mapped to the same resource with $\activityToSchedule$, i.e. $\mapping_l = \mapping_c$, the intervals in which $\activityToSchedule$ is scheduled are taken out of $D_l$. Secondly, for each successor $\activity_l$ of activity $\activityToSchedule$ the intervals are changed as $D_l^j = D_l^j \setminus \{[0,\scheduleHeur_j +  \executionTime_c - 1]\}$, since a successor can never start before a predecessor is completed. Next, the feasible solution is added to the \emph{set of already scheduled activities, represented by their schedules} $\ScheduledSet$, and $\queueToSchedule$ is updated to contain previously unscheduled activities, if there is any. If the current activity $\activityToSchedule$ is not schedulable, at least one activity has to be unscheduled. The activity to be unscheduled $\activityToUnschedule$ is found according to the rule $Un$ and this activity with all its successors are taken out of $\ScheduledSet$ (line~15). Next, the set of intervals $D$ is updated in the inverse manner compared to the previously described new activity insertion. To prevent cyclic scheduling and unscheduling of the same set of activities, a set $\RootProblems$ of \emph{activities that were problematic to schedule} is maintained. Therefore, the activity to schedule $\activityToSchedule$ has to be added to $\RootProblems$ (line~17) if it is not there yet. 
\begin{algorithm}[ht]

\caption{3-Level Scheduling Heuristic}
\label{alg:Heuristic}
\LinesNumbered
\KwIn{$\activitySet$}
$\ScheduledSet~=~\emptyset$, $\RootProblems~=~\emptyset$, $\ScheduledFromScratch~=~\emptyset$\;
$D$.initialize()\;
$\queueToSchedule$ = sort($\activitySet$, Pr)\;
\While{$|\ScheduledSet| < |\activitySet|$}{
	$\activityToSchedule$ = $\queueToSchedule$.pop()\;
	\textcolor{olive}{\textls[100]{{\small// Schedule $\activityToSchedule$ alone} }} \\
	$\scheduleHeur$ = SubModel($\activityToSchedule$, NULL, $D$)\;
	\eIf{SubModel found feasible solution}{  
		\textcolor{olive}{\textls[100]{{\small// Add previously unscheduled activities to $\queueToSchedule$} }} \\
		$\queueToSchedule$.update()\;
		$\ScheduledSet$.add($\scheduleHeur$) \tcp*{First scheduling level}
		$D$.update()\;
	}{
	 	$\activityToUnschedule$ = getActivityToUnschedule($\ScheduledSet$, Un)\;
	 	%\textcolor{olive}{\textls[100]{{\small// Remove $\activityToUnschedule$ and successors from $\ScheduledSet$} }} \\
		\textls[0]{$\ScheduledSet$ = $\ScheduledSet \setminus \{\activityToUnschedule \cup \activityToUnschedule.suc\}$}\;
		$D$.update()\;
		$\RootProblems$.add($\activityToSchedule$)\; %\: \: \: \: \: \: \: \: \: \: \: \textcolor{olive}{\textls[100]{// {\small$\activityToSchedule$ is problematic}}} \\
		\If{$\RootProblems$.contains($\activityToUnschedule$)}{ 
		\textcolor{olive}{\textls[100]{{\small// Schedule $\activityToSchedule$ and $\activityToUnschedule$ simultaneously} }} \\
			$\scheduleHeur$ = SubModel($\activityToSchedule$, $\activityToUnschedule$, $D$) \; 
			\eIf{SubModel found feasible solution}{
				$\ScheduledSet$.add($\scheduleHeur$) \tcp*{Second level}
			}{
				\textcolor{olive}{\textls[100]{{\small// Leave in $\ScheduledSet$ only activities from $ \ScheduledFromScratch$ and predecessors of $\activityToSchedule$ and $\activityToUnschedule$} }} \\
				$\ScheduledSet = \ScheduledFromScratch \cup \activityToSchedule.pr \cup \activityToUnschedule.pr$\;
				$D$.update()\;
				$\scheduleHeur$ = SubModel($\activityToSchedule$, $\activityToUnschedule$, $D$)\;
				\eIf{SubModel found feasible solution}{
					$\ScheduledSet$.add($\scheduleHeur$) \tcp*{Third level}
					$D$.update()\;
					$\ScheduledFromScratch$.add($\activityToSchedule$, $\activityToUnschedule$, $\activityToSchedule.pr$, $\activityToUnschedule.pr$)\;
				}{
					\KwOut{FAIL}
					}
			}
		}{}
	}
}
\KwOut{$\ScheduledSet$}

\end{algorithm}
In case the activity to be unscheduled is not problematic, i.e. $\activityToUnschedule \not\in \RootProblems$, the algorithm schedules $\activityToSchedule$ without $\activityToUnschedule$ scheduled in the next iteration. Otherwise, the second level of scheduling  takes place, as shown in Figure~\ref{fig:heuristic_description}. In this case, the sub-model is called to schedule $\activityToSchedule$ and $\activityToUnschedule$ simultaneously (line~20) and the set of two schedules $\scheduleHeur$ are added to $\ScheduledSet$. 

Sometimes, even simultaneous scheduling of two problematic activities does not help and a feasible solution does not exist with the given set of previously scheduled activities $\ScheduledSet$. If this is the case, we go to the third level of scheduling and try to schedule these two activities almost from scratch, leaving in the set of scheduled activities $\ScheduledSet$ only the set $\ScheduledFromScratch$ of activities \emph{that were previously scheduled in level 3} and the predecessors of $\activityToSchedule$ and $\activityToUnschedule$ (line~29). The set $\ScheduledFromScratch$ is introduced to avoid the situation where the same pair of activities is scheduled almost from scratch more than once, which is essential to guarantee termination of the algorithm. At the third scheduling level, the algorithm runs the sub-model to schedule $\activityToSchedule$ and $\activityToUnschedule$ with a smaller set of scheduled activities $\ScheduledSet$. In case of success, the obtained schedules $\scheduleHeur$ are added to $\ScheduledSet$ (line~29) and $\activityToSchedule$ together with $\activityToUnschedule$ and their predecessors $\activityToSchedule.pr$ and $\activityToUnschedule.pr$ are added to the set of activities $\ScheduledFromScratch$, scheduled almost from scratch. If the solution is not found at this stage, the heuristics fails to solve the problem. Thus, the 3-LS heuristic proceeds iteration by iteration until either all activities from $\activitySet$ are scheduled or the heuristic algorithm fails. Note that the same structure of the algorithm holds for both ZJ and JC cases. 

%{\color{red} Note that a possible extension of the algorithm while solving mapping and routing problems with fully-switched networks could be running the presented algorithm in a loop for a given mapping and routing, finding activities that cause infeasible solution and changing mapping and routing correspondingly to improve the solution.}

\subsection{Priority and Unscheduling Rules}
\label{rules}

There are two rules in the 3-LS heuristic: $Pr$ \emph{to set the priority} of insertion and $Un$ \emph{to select the activity to unschedule}. The rule to set the priorities considers information about activity periods $\periodSet$, activity execution times $\executionTimeSet$, the critical lengths of the predecessors execution before $\criticalTimeBefore$ and after $\criticalTimeAfter$ and the jitter requirements $\jitter$. However, not only the jitter requirements of the activity need to be considered, but also the jitter requirements of its successors. The reason is that if some non-jitter-critical activity would precede an activity with a critical jitter requirement in the dependency graph, the non-jitter-critical activity postpones the scheduling of the jitter-critical activity, resulting in the jitter-critical activity not being schedulable. We call this parameter \emph{inherited jitter} of an activity, computed as $\promotedJitter_i = \min_{\activity_j \in Pred_i} {\jitter_j}$. Using the inherited jitter for setting the priority is similar to the concept of priority inheritance~\cite{sha1990priority} in event-triggered scheduling. 

Thus, the priority assignment scheme $Pr$ sets the priority of each activity $\activity_i$ to be a vector of two components $priority_{sched} = (\min(\lengthOfSchedInterval_i,\promotedJitter), \max(\lengthOfSchedInterval_i,\promotedJitter))$, where $\lengthOfSchedInterval_i$ is the worst-case slack of the corresponding DAG, defined in Equation~\eqref{opt1}. The priority is defined according to lexicographical order, i.e. by comparing the first value in the vector and breaking the ties by the second. We compare first by the most critical parameter, either jitter $\promotedJitter_i$ or the worst-case slack $\lengthOfSchedInterval_i$, since those two parameters reflect how much freedom the activity has to be scheduled and the activity with less freedom should be scheduled earlier. This priority assignment strategy considers all of the aforementioned parameters, by definition outperforming the strategies that compare based on only subsets of these parameters.

The rule $Un$ to choose the activity to unschedule is a multi-level decision process. The general rules are that \emph{only activities that are mapped to the resource where activity $\activityToSchedule$ is mapped are considered} and we do not unschedule the predecessors of $\activityToSchedule$. Moreover, the intuition behind the $Un$ rule is that unscheduling activities with very critical jitter requirements or with already scheduled successors should be done only if no other options exist. The exact threshold for being very jitter-critical depends on the size of the problem, but based on experimental results we set the threshold of a high jitter-criticality level to the minimum value among all periods. Thus, whether or not an activity is very jitter-critical is decided by comparing its jitter to the threshold value $thresh = \min_{\activity_i \in \activitySet} \period_{i}$. 

The rule $Un$ can hence be described by three steps that are executed in the given order:
\begin{enumerate}
\item If there are activities without already scheduled successors and with $\jitter_i \geq thresh$, choose the one with the highest $\lengthOfSchedInterval_i$.
\item If all activities have successors already scheduled, but activities with $\jitter_i \geq thresh$ exist, we choose the one according to the vector $priority_{unsched} = (\text{number }$ $\text{of successors}$ $\text{scheduled}, \lengthOfSchedInterval_i)$ comparing lexicographically.
\item Finally, if all activities have $\jitter < thresh$, the step chooses the activity to unschedule according to the priority vector $priority_{unsched} = (\promotedJitter, \lengthOfSchedInterval_i)$ comparing lexicographically.
\end{enumerate}
Step~1 is based on the observation that activities with very critical jitter requirements are typically hard to schedule, unlike those with no jitter requirements or less critical ones. Besides, unscheduling many activities instead of one may cause prolongation of the scheduling process and possibly more complications with further scheduling of successors. Moreover, since only activities of cause-effect chains are a part of precedence relations, there are many activities with no predecessors and successors that can be unscheduled. This is typical for the automotive domain~\cite{WATERSAutomotive2015}. Step~2 allows unscheduling of activities with already scheduled predecessors, preferring to keep in the schedule activities with critical jitter requirements. Step~3 states that if all of the activities have very critical jitter requirements, the activity with the highest value of inherited jitter should be unscheduled. In all three steps, ties are broken by choosing the activity with higher worst-case slack $\lengthOfSchedInterval$ value by the same intuition as in the $Pr$ rule.

We have experimentally determined that comparing to the unscheduling rule with only worst-case slack ($\lengthOfSchedInterval_i$) considered, the gain of the presented unscheduling rule is 5\% more utilization achieved on average.

\section{Experiments}
\label{experiment}

This section experimentally evaluates and compares the proposed optimal models and 3-LS heuristic on synthetic problems with jitter requirements set differently to show the benefits of JC scheduling in terms of utilization. Furthermore, we quantify the trade-off of additional cost in terms of memory to store the schedule and increase in computation time versus this gained utilization. Note that the goal of this section is to show the advantages and disadvantages of the JC approach. The experimental setup is presented first, followed by experiments that evaluates the proposed exact and heuristic approaches for different jitter and period requirements. We conclude by demonstrating our approach on a case study of an Engine Management System with more than 10000 activities to be scheduled.

\subsection{Experimental Setup}

Experiments are performed on problem instances that are generated by a tool developed by Bosch~\cite{WATERSAutomotive2015}. There are five sets of 100 problem instances, each set containing 20, 30, 50, 100 and 500~tasks, respectively. The same problem instance is presented with different jitter requirements. The generation parameters for each dataset are presented in Table~\ref{table:model_generation}, and the granularity of the timer is set to be $1$~$\mu$s. Message communication times are computed for the considered platform with the following parameters: bandwidth $bnd =  400$~MB/s and latency $lat = 50$ clock cycles. 
%, where the added latency has no impact on the resulting communication times due to the chosen granularity.

The mapping is found as described in Section~\ref{mapping} so that load is balanced across the cores, i.e. the resulting mapping utilizes all cores approximately equally. The resulting problem instances contain 30-45, 50-65, 90-130, 180-250 and 1500-2000~activities (tasks and messages) for sets with 20, 30, 50, 100 and 500~tasks, respectively. 

While we initially assume a system with 3~cores connected over a crossbar (resulting in 6 resources), inspired by the Infineon Aurix Tricore Family TC27xT, the approach can scale to a higher number of cores, as shown in Section~\ref{difArchtsExp}. 

The metric for the experiments on the synthetic datasets is the maximum utilization for which the problem instance is still schedulable. The utilization is defined as $\utilization_y = \sum_{\activity_i \in \activitySet: \mapping_i = y}{\frac{\executionTime_i}{\period_i}}$ on each resource $y = 1,\cdots,6$. To achieve the desired utilization on each resource, the execution times of activities are scaled appropriately. The experiments always start from a utilization of 10\%, increasing in steps of 1\%, solving until the approach is not able to find a feasible solution. The last utilization value for which the solution was found is set as the \emph{maximum utilization} of the approach on the problem instance. Although this approach to set the maximum schedulable utilization may not be completely fair, the utilization is monotonic in most cases. Therefore, we have chosen to approximate the results by setting this rule to get results that are easier to interpret. 

\begin{table}
\caption{Generator parameters for the sets of problem instances}
\label{table:model_generation} 
\centering
\begin{tabular}{|c|c|c|c|c|}
\hline
\emph{Set} & $|\taskSet|$ & \emph{$\periodSet$ [ms]} & \emph{Variable accesses} & \emph{Chains } \\
&&& \emph{per task} & \emph{per task}\\ 
\hline 
\rowcolor{LightGray}
1 & 20  & $1,2,5,10$ & 4  & 4   \TBstrut \\ \hline
2 & 30   & $1,2,5,10$ & 4  & 6    \TBstrut\\ \hline
\rowcolor{LightGray}
3 & 50  & $1,2,5,10,20,50,100$ & 4   & 8  \TBstrut\\ \hline
4 & 100  & $1,2,5,10,20,50,100$ & 4 & 15 \TBstrut \\ \hline
\rowcolor{LightGray}
5 & 500 & $1,2,5,10,20,50,100$ & 8 & 50 \TBstrut \\ \hline
\end{tabular}
\end{table}

Experiments were executed on a local machine equipped with Intel Core i7~(1.7~GHz) and 8~GB memory. The ILP model and ILP part of the 3-LS heuristic were implemented in IBM ILOG CPLEX Optimization Studio~12.5.1 and solved with the CPLEX solver using concert technology, while the SMT model was implemented in Z3~4.5.0. The ILP, SMT and heuristic approaches were implemented in the JAVA programming language.

 \subsection{Results} 
First, the experiments compare the computation time of the two optimal ILP and SMT approaches to show for which problem instances it is advantageous to use each of the approaches. Secondly, we evaluate trade-off between the maximum achievable utilization and computation time of the 3-LS heuristic and the optimal approaches for differently relaxed jitter requirements. Thirdly, since memory consumption to store the final schedule is also a concern, the trade-off between solution quality and required memory is evaluated for systems of different sizes. Finally, a comparison of different period settings is presented to show the applicability of the approach to different application domains and to evaluate the behavior of both ZJ and JC approaches for periods set differently. A time limit of 3~000~seconds per problem instance was set for the optimal approaches to obtain the results in reasonable time. Note that the best solution found so far is used if the time limit is hit.

\subsubsection{Comparison of the ILP and SMT models with different jitter requirements}
%Figure~\ref{fig:difJitOptCompTime1} 
%and~\ref{fig:difJitOptCompTime2} 
First of all, we compare the computation time distribution for Set~1 and Set~2 
(of smaller instance sizes with 30-45 activities and 50-65 activities, respectively) for SMT and ILP approaches with jitter requirements of each activity $\activity_i \in \activitySet$ set to $\jitter_i = \frac{\period_i}{2}$, $\jitter_i = \frac{\period_i}{5}$, $\jitter_i = \frac{\period_i}{10}$ and $\jitter_i = 0$. Since the first problem instance from Set~3  was computing for two days before it was stopped with no optimal solution found for both SMT and ILP models, the experiments with optimal approaches only use the first two sets. We will return to the larger sets in Section~\ref{heurExp} when evaluating the 3-LS heuristic. 

The distribution is shown in the form of box plots~\cite{kampstra2008beanplot}, where the quartile, median and three quartiles together with outliers (plus signs) are shown. Outliers are numbers that lie outside $1.5 \times$the interquartile range away from the top or bottom of the box that are represented by the top and the bottom whiskers, respectively. Note that outliers were also successfully solved within the time limit. 
\begin{figure}[h]
\label{compTimeOpt}
\centering
\epsfig{file=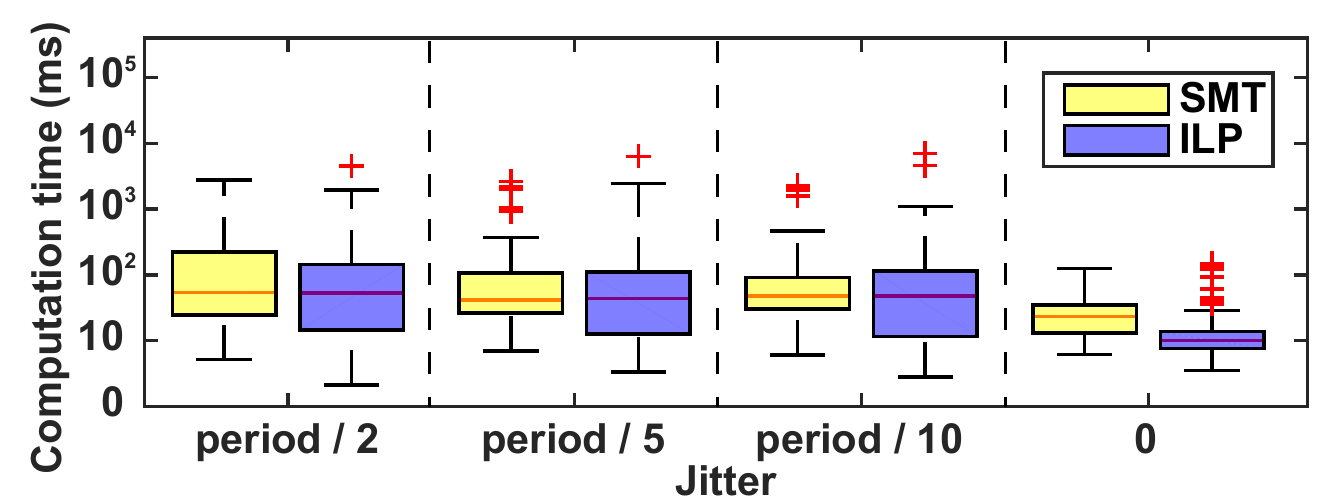,width=1.0\columnwidth}
\caption{Computation time distribution for the SMT and ILP models with different jitter requirements for Set~1.}
\label{fig:difJitOptCompTime1}
\end{figure}

The number of problem instances from Set~1 and Set~2, that the optimal approaches failed to solve within the given time limit is shown in Table~\ref{tab:failed}. Moreover, Figure~\ref{fig:difJitOptCompTime1} displays the computation time distribution on Set~1, where only problem instances, that both the ILP and SMT solvers were able to optimally solve all jitter requirements within the timeout period are included. For Set~1, it is 82 (out of 100), and for Set~2, it is 21 (out of 100) problem instances. The computation time distribution for Set~2 in show a similar trend, but since the sample is too small to be representative, we do not display them.
%
%\begin{figure}[h]
%\centering
%\epsfig{file=figures/comparison_dif_jitters_time_set2.pdf,width=1.0\columnwidth}
%\caption{Computation time distribution for the SMT and ILP models with different jitter requirements for Set~2.}
%\label{fig:difJitOptCompTime2}
%\end{figure}
%
\begin{table}[]
\centering
\caption{Number of problem instances optimal approaches failed to solve before to time limit of 3~000 seconds}
\label{tab:failed}
\begin{tabular}{|l|l|l|l|l|l|l|l|l|}
\hline
\multirow{2}{*}{$\jitter_i$} & \multicolumn{2}{c|}{$\period_i/2$}                      & \multicolumn{2}{c|}{$\period_i/5$}                     & \multicolumn{2}{c|}{$\period_i/10$}                     & \multicolumn{2}{c|}{$0$}                         \TBstrut       \\ \cline{2-9} 
                             & \multicolumn{1}{c|}{Set~1} & \multicolumn{1}{c|}{Set~2} & \multicolumn{1}{c|}{Set1} & \multicolumn{1}{c|}{Set~2} & \multicolumn{1}{c|}{Set~1} & \multicolumn{1}{c|}{Set~2} & \multicolumn{1}{c|}{Set~1} & \multicolumn{1}{c|}{Set~2} \TBstrut \\ \hline
ILP                          & 14                         & 76                         & 9                         & 53                         & 6                          & 45                         & 4                          & 27                    \TBstrut      \\ \hline
SMT                          & 2                          & 51                         & 3                         & 13                         & 2                          & 9                          & 2                          & 7                     \TBstrut      \\ \hline
\end{tabular}
\end{table}

The results in Table~\ref{tab:failed} show that for more difficult problem instances the SMT model is significantly better than the ILP model in terms of computation time, since it is able to solve more problem instances within the given time limit. On the other hand, the comparison on the problem instances that both approaches were able to solve in Figures~\ref{fig:difJitOptCompTime1} indicates that the ILP runs faster on simpler problem instances that can be found at the bottom of the boxplots in Figure~\ref{compTimeOpt}. 
%Moreover, when both models are able to solve a problem instance, the results are the following. On Set~1, the average computation times of the two approaches does not differ more than by 0.2 seconds for all jitter requirements, being maximally 2.7 seconds for $\jitter_i = \frac{\period_i}{2}$ and minimally 0.2 seconds for $\jitter_i = 0$. For Set~2, the ILP model runs twice as long as SMT model on average (15 seconds versus 7.3) on the problem instances with $\jitter_i = \frac{\period_i}{2}$. %,while with ZJ requirements the difference in computation time becomes 0.5 seconds and 0.7 seconds for ILP and SMT model, respectively, for.
As one can see, more relaxed jitter requirements result in longer computation time, which is a logical consequence of having larger solution space.

Thus, \emph{the SMT model is more efficient than the ILP model for the considered problem on more difficult problem instances, while the ILP model shows better results for simpler instances, which justifies the usage of the ILP model in the 3-LS heuristic. Besides, more relaxed jitter requirements cause longer computation time for the optimal approaches}. Therefore, the SMT approach results are used for further comparison with the 3-LS heuristic.

\subsubsection{Comparison of the optimal and heuristic solutions with different jitter requirements}
\begin{figure}[h]
\centering
\epsfig{file=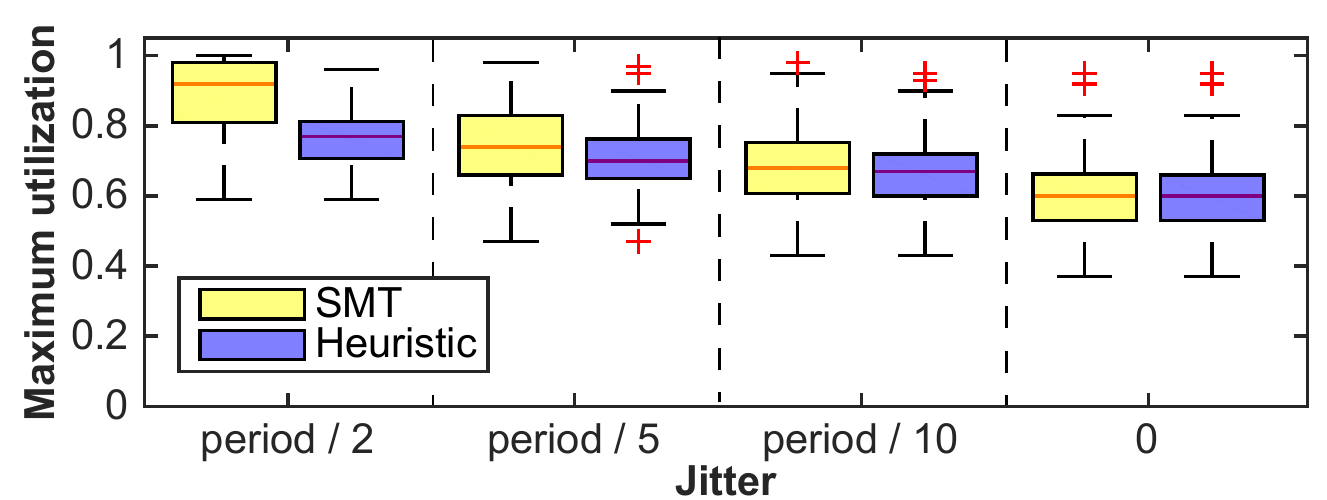,width=1.0\columnwidth}
\caption{Maximum utilization distribution for the optimal SMT and 3-LS heuristic approaches with different jitter requirements for Set~1.}
\label{fig:comparisonWithDifferentJitter1}
\end{figure}

Figure~\ref{fig:comparisonWithDifferentJitter1} shows the distribution of the maximum utilization on Set~1 for SMT and 3-LS heuristic with different jitter requirements. For comparison, we use the solution with the highest utilization, while the low value of initial utilization guarantees that at least some solution is found. The time limit caused 3 problem instances in Set~1 not to finish when using the SMT approach and these instances are not included in the results. The results for Set~2 are similar to that of Set~1, but due to small number of solvable instances we do not show them.
%
%\begin{figure}[h]
%\centering
%\epsfig{file=figures/comparison_dif_jitters_crit_set2.pdf,width=1.0\columnwidth}
%\caption{Maximum utilization distribution for the optimal and 3-LS heuristic approaches with different jitter requirements for Set~2.}
%\label{fig:comparisonWithDifferentJitter2}
%\end{figure}
%
The results for the optimal approach shows that stricter jitter requirements cause lower maximum achievable utilization. Namely, the average maximum utilization is 89\%, 75\%, 69\%, 61\% for Set~1 and 95\%, 81\%, 74\%, 67\% for Set~2 for the instances with jitter requirements equal to half, fifth, tenth of a period and zero, respectively. Meanwhile, the comparison of the 3-LS heuristic to the optimal solution reveals that the average difference goes from 17\% and 23\% for Set~1 and Set~2, respectively, with the most relaxed $\jitter_i = \frac{\period_i}{2}$ to 0.1\% for both sets with ZJ scheduling. This difference for problem instances with more relaxed jitter requirements is caused by very large complexity of the problem solved. Furthermore, while the heuristic solves all problem instances in hundreds of milliseconds, the SMT model fails on 62 problem instances out of 200 within a time limit of 3~000 seconds. This reduction of the computation time by the heuristic is particularly important during design-space exploration, where many different mappings or platform instances have to be considered. In that case, it is not possible to spend too much time per solution. 

Hence, we conclude that \emph{heuristic performs better with tighter jitter requirements and hence particularly well for ZJ scheduling, resulting in an average degradation of 7\% for all instances. Moreover, unlike the SMT model, the 3-LS heuristic always finds feasible solutions in hundreds of milliseconds, hence providing a reasonable trade-off between computation time and solution quality}. 

%\emph{optimal JC scheduling takes more time, but achieves higher utilization than optimal ZJ scheduling}. 

%One reason for the 3-LS heuristic to fail is that it is not flexible enough in assigning precise start times, i.e. it is not able to slightly move the already scheduled jobs of activities so that the currently scheduled activity still have a chance to be scheduled. Thus, an efficient improvement can be a more flexible solution representation that will allow the existing schedule to adjust to the new activities to be scheduled. We leave this improvement as future work. 

\subsubsection{Comparison of the heuristic with ZJ and JC scheduling}
\label{heurExp}
\begin{figure}[h]
\centering
\epsfig{file=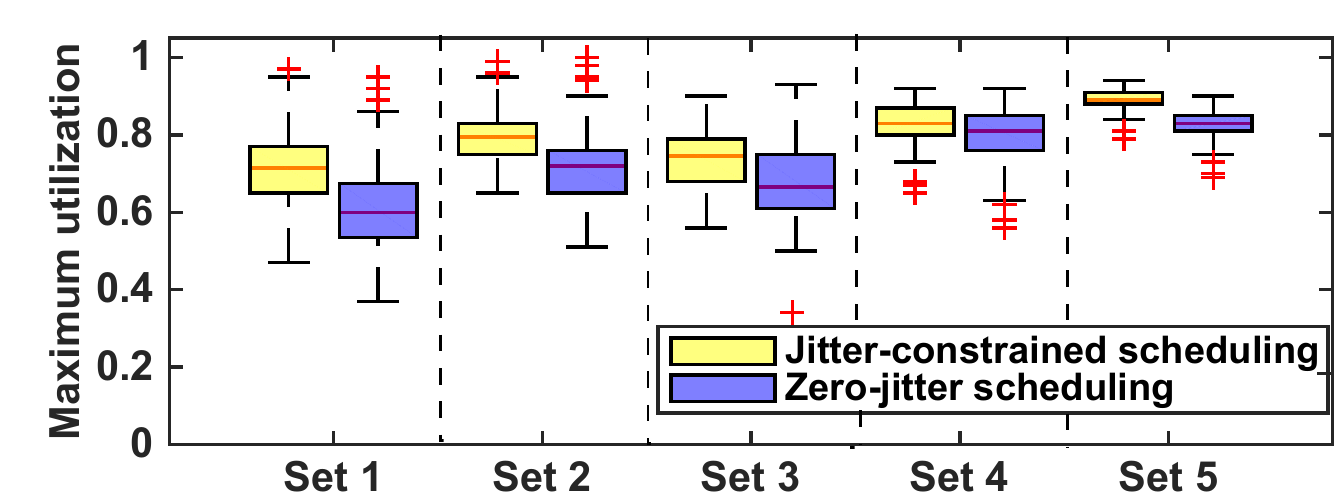,width=1\columnwidth}
\caption{Maximum utilization distribution for the 3-LS heuristic with jitter-constrained and zero-jitter requirements in sets with 20, 30, 50, 100 and 500 tasks.}
\label{fig:heurToHeurComparison}
\end{figure}

While the previous experiment focused on comparing the optimal approach and the heuristic, therefore using only smaller problem instances, this experiment evaluates the 3-LS heuristic on all sets. Due to time restrictions, only two jitter requirements were considered, $\jitter_i = \frac{\period_i}{5}$ and $\jitter_i = 0$. Figure~\ref{fig:heurToHeurComparison} shows the distribution of the maximum utilization for the 3-LS heuristic on Sets~1 to~5. In all sets, 100 problem instances were used for this graph. The results show that with growing size of the problem instance, the maximum utilization generally increases. The average difference in maximum utilization of the 3-LS heuristic on the problem instances with JC and ZJ requirements is 15.3\%, 9.7\%, 8.6\%, 4.2\% and 7.5\% for Sets~1 to~5, respectively, with JC achieving higher utilization. The decreasing difference with growing sizes of the problem is caused by the growing average utilization. For instance, the average maximum utilization for Set 5 is 89.1\% for the problem instances with JC requirements and 82.6\% for the problem instances with ZJ requirements, pushing how far the maximum utilization for the JC scheduling can go. This tendency of increasing maximum utilization for the ZJ scheduling can be intuitively supported by the fact that more and more activities are harmonic with each other, which results in easier scheduling. 
%However, these synthetic instances are created to be almost harmonic to simplify scheduling process. 
In reality, harmonization costs a significant amount of over-utilization, especially when activities with smaller periods are concerned. On problem instances without harmonized activity periods the JC scheduling can show notably better results for larger instances compared to ZJ scheduling, as shown in Section~\ref{expDifPer}.

%\vspace{-0.5cm}
\begin{figure}[h]
\centering
\epsfig{file=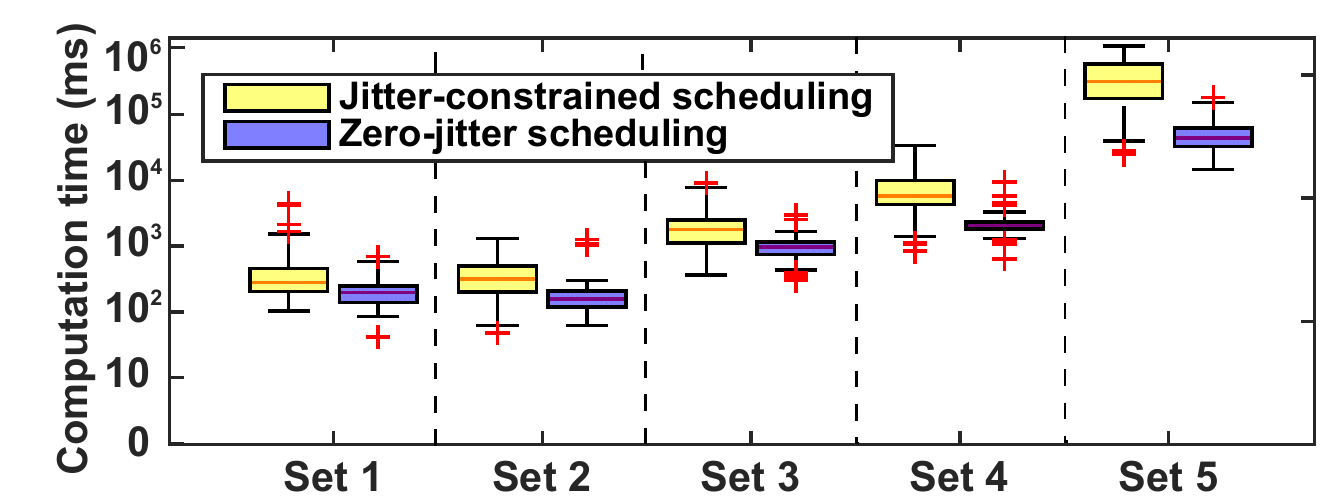,width=1\columnwidth}
\caption{Computation time distribution for the 3-LS heuristic with jitter-constrained and zero-jitter requirements in sets with 20, 30, 50, 100 and 500 tasks.}
\label{fig:heurToHeurComparisonCompTime}
\end{figure}

Figure~\ref{fig:heurToHeurComparisonCompTime} shows the computation time of ZJ and JC using the 3-LS heuristic. Similarly to the optimal approach, the 3-LS heuristic takes longer to solve problem instances with JC requirements due to the larger solution space. Specifically, the average computation time for JC heuristic for Sets~1 to~5 are 0.3, 0.6, 3.6, 14.5 and 1003.6 seconds, respectively, while for ZJ scheduling it is 0.15, 0.28, 1.6, 4 and 109 seconds. Thus, solving a problem instance with JC requirements with 1500-2000 activities takes less than 17 minutes on average, which is still reasonable. Hence, \emph{the 3-LS heuristic with JC scheduling provides better results, but requires more time than the 3-LS heuristic with ZJ scheduling}.

To summarize this experiment, JC scheduling \emph{is promising in terms of maximum utilization, as it schedules with up to 55\% higher resource utilization}. Besides, the computation time of the proposed heuristic is affordable even for larger problem instances, while the optimal models fail to finish in reasonable time already for much smaller instances. Moreover, \emph{the proposed heuristic solves the problem instances with ZJ requirements near-optimally with a difference of 0.1\% in schedulable utilization on average}. Generally, \emph{the JC heuristic provides more efficient solutions than the ZJ heuristic, while requiring longer computation time}.

\subsubsection{Evaluation of required memory and maximum utilization trade-off with different number of cores}
\label{difArchtsExp}
\begin{figure}[h]
\centering
\epsfig{file=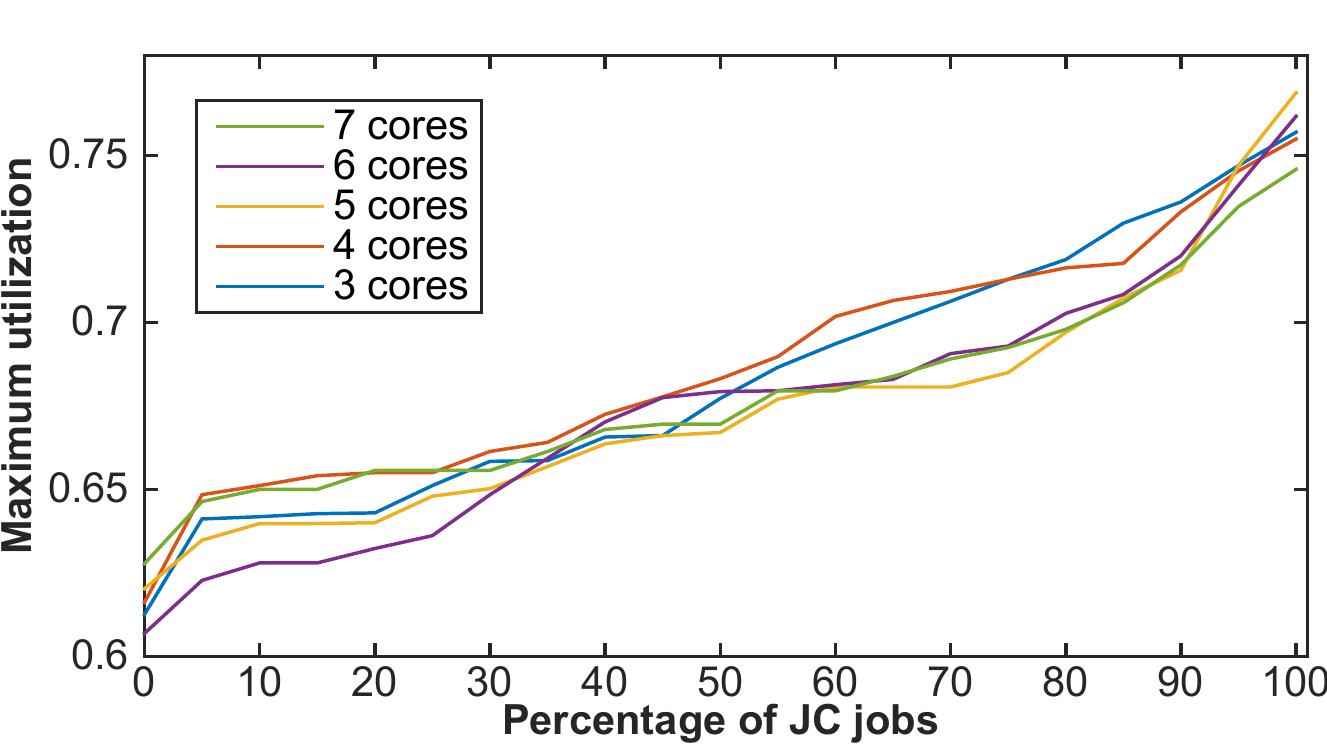,width=1.0\columnwidth}
\caption{Utilization distribution for different percents of JC activities for different architectures.}
\label{fig:difPercJittersComparison}
\end{figure}

The trade-off between maximum achievable utilization and the amount of memory required to store the schedule is evaluated by this experiment. Figure~\ref{fig:difPercJittersComparison} shows the average maximum utilization achieved on systems with different number of cores and with gradually increasing percentage of JC jobs on 50 problem instances from Set~2 (due to time restrictions). The jitter constraint is set to $\jitter_i = \frac{\period_i}{5}$ and the instances are solved to optimality. Furthermore, the problem instances with different numbers of cores are solved in steps of 5\% of jobs with zero-jitter requirement, which reflects how much memory is necessary to store the schedule for such solutions. Note that the execution times of the activities are scaled proportionally to the number of cores so that each resource has a required utilization. 

The results show that introducing more JC jobs and thus increasing memory requirements for storing the final schedules can significantly improve the average maximum utilization. Namely, for the architecture with 4 cores, the maximum utilization with all ZJ jobs is 61\%, while relaxing jitter requirements of half the jobs results in 69\% utilized resources, and relaxation all of the jobs increases the maximum utilization to 76\%. Concerning the required memory to store the schedule, the problem instances with 4 cores on average contain 80 jobs with JC scheduling and 49 jobs while scheduling in zero-jitter manner. Thus, assuming we need 8 bytes to store the schedule of one job, the memory overhead of relaxing jitter is 31 * 8 = 248 bytes, which is a reasonable price to pay for utilization gain of 15\% on average on each resource.  

Concerning the increasing number of cores, the results demonstrate that on average there is no significant dependency on how much cores we have in the system. Hence, \emph{JC scheduling can result in high utilization gain, although at the cost of increased memory requirements to store the resulting schedule}.

\subsubsection{Comparison of the different period settings}
\label{expDifPer}
\begin{figure}[h]
\centering
\epsfig{file=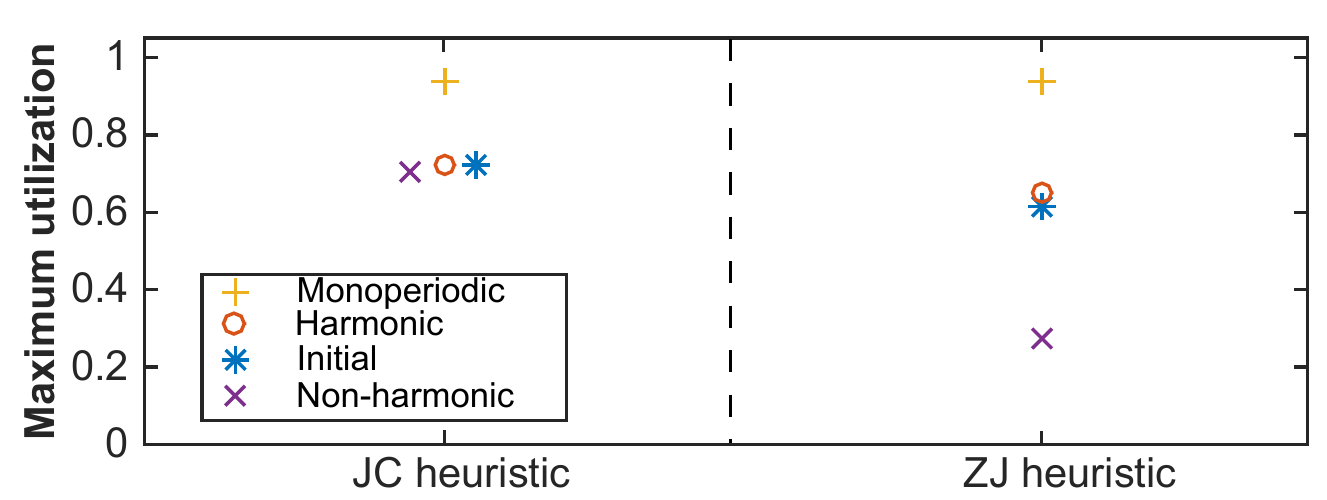,width=1\columnwidth}
\caption{Utilization distribution for problem instances with different periods.}
\label{fig:difPeriodsComparison}
\end{figure}

To show that the approach is applicable to other domains, an experiment with different period settings is performed. All problem instances from Set~2 are solved monoperiodically ($\period_i = 10$~ms for each activity $\activity_i \in \activitySet$), or with harmonic periods (activities with $\period_i = 2$~ms are changed to $\period_i = 5$~ms), or with initial periods (i.e. with periods 1, 2, 5, 10~ms), or with non-harmonic periods (with periods 2, 5, 7, 12~ms). Figure~\ref{fig:difPeriodsComparison} displays the average maximum utilization achieved by the 3-LS heuristic with ZJ and JC scheduling ($\jitter_i = \frac{\period_i}{5}$) on 100 problem instances from Set~2. Since the optimal approach was not able to solve 7 out of 10 first instances with non-harmonic periods within the given time limit, due to its complexity and extended hyper-period, the optimal approach results are not included in the figure.

The results show that the maximum utilization for both ZJ and JC is achieved when scheduling monoperiodically, which is explained by having less possible collisions in the resulting schedule. An interesting observation is that for JC scheduling all other period settings on average resulted in very similar maximum utilization, while the ZJ approach shows the variation of 27\% with non-harmonic periods, 62\% with initial periods and 65\% with harmonic period set. Relative insensitivity of JC scheduling to period variations can be caused by significantly larger solution space due to relaxation of strict jitter constraints. This allows to find solutions with high utilization even with the non-harmonic period setting. 

Besides, the same order of computation time distribution is shown by different period settings, i.e. monoperiodic scheduling is the fastest, while the problem instances in the non-harmonic period set result in the longest computation time. 

Thus, \emph{the proposed approach is applicable to other domains, where the application periods have different degree of harmonicity. Furthermore, increasing harmonicity of the period set results in higher maximum utilization, lower computation time and lower gain of JC scheduling in comparison with ZJ scheduling in terms of maximum utilization}.

\subsection{Engine Management System Case Study}
\label{use-case}
We demonstrate the applicability of the proposed 3-LS heuristic on an Engine Management System (EMS). This system is responsible for controlling the time and amount of air and fuel injected by the engine by considering the values read by numerous sensors in the car (throttle position, mass air flow, temperature, crankshaft position, etc). By design, it is one of the most sophisticated engine control units in a car consisting of 1000-2000 tightly coupled tasks that interact over 20000 to 30000 variables, depending on the features in that particular variant. A detailed characterization of such an application is presented by Bosch in~\cite{WATERSAutomotive2015}, along with a problem instance generator that creates input EMS models in conformance with the characterization.  

We consider such a generated EMS problem instance, comprising 2000~tasks with periods 1, 2, 5, 10, 20, 50, 100, 200 and 1000~ms and with 30000~variables in total, where each task accesses up to 12 variables. There are 60 cause-effect chains in the problem instance with up to 11 tasks in each chain. We consider the target platform to be similar to an Infineon AURIX Family TC27xT with a processor frequency of 125~MHz and an on-chip crossbar switch with a 16~bit data bus running at 200~MHz, thus having a bandwidth of 16-bit x 200~MHz / 8 = 400~MB/s. The time granularity is 1~$\mu$s, and the resulting hyper-period is 1000~ms. However, setting the hyper-period to be 100~ms results in a utilization loss of less than 0.5\%, arising from shortening the scheduling periods of tasks with periods 200~ms and 1000~ms and over-sampling, which is a reasonable sacrifice to decrease the memory requirements of the schedule. The tool in~\cite{WATERSAutomotive2015} provides the number of instructions necessary to execute each task, which is used to compute the worst-case execution time with the assumption that each instruction takes 3 clock cycles on average (including memory accesses that hit/miss in local caches).

The mapping of tasks to cores by the simple ILP described in Section~\ref{description} requires minimally 3~cores with the utilization approximately 89.6\% on each core and approximately 30\% on each input port of the crossbar. Moreover, the resulting scheduling problem has 10614~activities with 104721~jobs for the JC assumptions in total. Neither SMT nor ILP can solve this problem in reasonable time, but the JC heuristic with $\jitter_i = \frac{\period_i}{5}$ for all $\activity_i$ solves the problem in 43~minutes. By gradually introducing more activities $\activity_i$ with $\jitter_i = 0$, we have found a maximum value of 85\% ZJ activities for which the 3-LS heuristic is still able to find a solution, which takes approximately 12 hours. Note that the computation time has increased with introducing more ZJ activities due to more restricted solution space. However, to store the schedule in the memory for 0\% ZJ jobs, 104721 * 8 = 818~Kbytes of memory is required assuming that one job start time needs 8 bytes, while with 85\% ZJ jobs it is only 19394 * 8 = 152~Kbytes. \emph{Thus, for realistic applications the optimal approaches take too long, while the 3-LS heuristic approach is able to solve the problem in reasonable time. Moreover, increasing the percent of ZJ activities has shown to provide a trade-off between computation time and required memory to store the obtained schedule}.

\section{Conclusions}
\label{conclusions}
This article introduces a co-scheduling approach to find a time-triggered schedule of periodic tasks with hard real-time requirements that are executed on multiple cores and communicate over an interconnect. Moreover, precedence and jitter requirements are put on these tasks due to the nature of such applications in the automotive domain. To optimally solve the considered problem, we propose both an Integer Linear Programming (ILP) model and Satisfiability Modulo Theory (SMT) model with computation time improvements that exploit problem-specific information to reduce the computation time. Furthermore, a three-step heuristic scheduling approach, called 3-LS heuristic, where the schedule is found constructively is presented. The heuristic works in three levels, where the scheduling complexity and the time consumption grow for each level, providing a good balance between solution quality and computation time.

We experimentally evaluate the efficiency of the proposed optimal and heuristic approaches with jitter-constrained (JC) requirements, comparing to the widely used zero-jitter (ZJ) approach and quantify the gain in terms of maximum utilization of the resulting systems for the optimal and heuristic approaches. The results show that JC scheduling by the optimal approaches achieves higher utilization with an average difference of 28\% compared to optimal ZJ scheduling. Moreover, the experimental evaluations indicate that SMT model is able to solve more problem instances optimally within a given time limit than the ILP model, while the ILP model shows better computation time on simpler problem instances. We also show that the 3-LS heuristic solves the problem instances with ZJ requirements near-optimally. %, while the results for JC case show that there is the room for improvement, which is our plan for the near future. 
The computation time of the proposed heuristic is acceptable even for larger problem instances, while the optimal models fail to finish in reasonable time already for smaller problem instances. Furthermore, the approach is demonstrated on a case study of an Engine Management System, where 2000 tasks are executed on cores, sending around 8000 messages over the interconnect. Here, we show that for realistic applications, the proposed SMT solution takes too long and the 3-LS heuristic is able to find the solution in reasonable time, providing a trade-off between required memory to store the schedule and computation time depending on percent of activities with zero-jitter requirements.

\section*{Acknowledgments}

This work was supported by the European Union’s Horizon 2020 research and innovation programme under grant agreement No. 688860 (HERCULES), and Eaton European Innovation Centre.  It was also partially supported by National Funds through FCT/MEC (Portuguese Foundation for Science and Technology) and co-financed
by ERDF (European Regional Development Fund) under
the PT2020 Partnership, within project UID/CEC/04234/2013
(CISTER Research Centre); also by FCT/MEC and the EU
ARTEMIS JU within project(s) ARTEMIS/0001/2013 - JU
grant nr. 621429 (EMC2).

\begin{comment}

\begin{itemize}
\item Runnables in cause-effect chains with different activation patterns. Questions are how to define end-to-end latency and precedence requirements. Paper~\cite{feiertag2008compositional}.
\item Bound hyper-period. Many interesting issues.
\item Deadlines later than the end of the period.
\end{itemize}

\begin{itemize}
\item We assume that the activation pattern of all the runnables in any cause-effect chains equals to the minimum activation pattern of runnables in this cause-effect chains, assuming they are all the same.
\item We set end-to-end latency requirements to be equal to the period. The motivation is not to have ''old data''.
\item We use first bin packing algorithm to solve mapping problem (that is typically given by a problem description).
\item We assume that each instruction takes 5 cycles and compute execution time correspondingly.
\item The communication time is obtained from the size of the label that is shared by runnables.
\end{itemize}

\end{comment}

\bibliographystyle{abbrv}
\bibliography{bibliography}

%\bibliographystyle{template/IEEEtran}
%\bibliography{template/IEEEabrv,bibliography}
\end{document}